\documentclass[aip,amsmath,amssymb,reprint]{revtex4-1}

\usepackage{graphicx}
\usepackage{dcolumn}
\usepackage{bm}

\usepackage[utf8]{inputenc}
\usepackage[T1]{fontenc}
\usepackage{mathptmx}
\usepackage{etoolbox}

\makeatletter
\def\@email#1#2{%
 \endgroup
 \patchcmd{\titleblock@produce}
  {\frontmatter@RRAPformat}
  {\frontmatter@RRAPformat{\produce@RRAP{*#1\href{mailto:#2}{#2}}}\frontmatter@RRAPformat}
  {}{}
}%
\makeatother

\begin{document}


\title{Improved algorithm for a two-dimensional Darwin particle-in-cell code} 



\author{Dmytro Sydorenko}
\email[]{sydorenk@ualberta.ca}
\affiliation{University of Alberta, Edmonton, AB, Canada}

\author{Igor D. Kaganovich}
\affiliation{Princeton Plasma Physics Laboratory, Princeton, NJ}

\author{Alexander V. Khrabrov}
\affiliation{Princeton Plasma Physics Laboratory, Princeton, NJ}

\author{Stephane A. Ethier}
\affiliation{Princeton Plasma Physics Laboratory, Princeton, NJ}

\author{Jin Chen}
\affiliation{Princeton Plasma Physics Laboratory, Princeton, NJ}

\author{Salomon Janhunen}
\affiliation{Los Alamos National Laboratory, Los Alamos, NM}

\date{\today}

\begin{abstract}

A two-dimensional particle-in-cell code for simulation of low-frequency electromagnetic processes in laboratory plasmas has been developed. The code uses the Darwin method omitting the electromagnetic wave propagation. The Darwin method separates the electric field into solenoidal and irrotational parts. The irrotational electric field is the electrostatic field calculated with the direct implicit algorithm. The solenoidal electric field is calculated with a new algorithm based on the equation for the electric field vorticity. The new algorithm is faster and more reliable than the Streamlined Darwin Field Formulation introduced decades ago. The system of linear equations in the new algorithm is solved using a standard iterative method. The code is applied to simulate an inductively coupled plasma with the driving current flowing around the plasma region.

\end{abstract}


\maketitle 


\section{\label{sec:01} Introduction}

In laboratory plasma devices and in plasma material processing reactors, the plasma is often dense (density $10^{17}~\text{m}^{-3}$ and up), relatively cold (electron temperature a few eV), and has dimensions of tens of centimeters. In this case, the plasma size is thousands of electron Debye length in one dimension. If the neutral pressure is a few mTorr, the electron mean free path may exceed the size of the plasma. Kinetic effects in such plasmas are important and can be studied numerically using particle-in-cell (PIC) codes. Explicit PIC algorithms are simple and robust, but they must resolve the electron Debye length and the plasma frequency.\cite{BirdsallBook1991} Combination of the large number of cells in the numerical grid and the small time step makes the explicit algorithms numerically expensive, especially for simulations resolving more than one spatial dimension. Implicit PIC algorithms are free from these requirements and allow to reach the final state in simulation much sooner compared to explicit codes. On the other hand, fully implicit algorithms are complex and involve solution of systems of nonlinear equations.\cite{ChenJCP2011,AngusJCP2024} A simpler predictor-corrector approach is implemented in the direct implicit method.\cite{LangdonJCP1983} The direct implicit algorithm is not energy conserving, but effects of numerical heating or cooling may be minimized by proper selection of the cell size and time step.\cite{CohenJCP1989}

Electromagnetic simulations using explicit numerical schemes have a strong limitation on the simulation time step because they have to resolve waves propagating with the light speed.\cite{YeeIEEETAP1966} This limitation does not apply to fully implicit electromagnetic codes which also have to solve a system of nonlinear equations.\cite{MatteiJCP2017} Another way to bypass this limitation is to omit propagation of electromagnetic weaves and use a method based on the Darwin approach where the inductive (solenoidal) electric field is neglected in the Ampere-Maxwell  law.\cite{DarwinPM1920,NielsonMCP1976,LeeNIMPRA2005,EreminJPD2013,BarnesJCP2022}

The PIC code described below is developed based on the two-dimensional Darwin direct implicit PIC code DADIPIC of Gibbons and Hewett.\cite{GibbonsJCP1995} The electrostatic part of the algorithm is similar to DADIPIC (though it is not an exact copy), while the electromagnetic part has significant differences. DADIPIC uses the Streamlined Darwin Field (SDF) Formulation to find solenoidal components of the electric field.\cite{HewettJCP1992} An advantage of the SDF method is the simplicity of boundary conditions it uses. A drawback is that the method produces a system of linear equations which is difficult to solve using most standard iterative linear solvers. Previously, the SDF system of equations was solved using the dynamic alternating-directions implicit technique.\cite{HewettJCP1992,GibbonsJCP1995} In this paper, the solenoidal electric field is obtained from an equation for the field vorticity. The system of linear equations resulting from this approach is solved with the bi-conjugate gradient (KSPBCGSL) solver of the Portable Extensible Toolkit for Scientific Computation (PETSc)\cite{petsc-web-page} with a multigrid preconditioner (PCHYPRE) from the LLNL package HYPRE.\cite{hypre-web-page}

The code is applied to simulate a side-coil inductively coupled plasma (ICP) system. The simulated system has rather high neutral density and antenna current. Such values are selected to test code stability and performance under extreme conditions rather than to investigate kinetic effects. Nevertheless, the simulation demonstrates transition from the classical to anomalous skin effect as the plasma density grows and the skin layer size becomes comparable to the electron mean free path. Also, the electron velocity distribution function in the simulation is non-Maxwellian.

The paper is organized as follows. Section~\ref{sec:02} contains description of the algorithm. Simulation parameters and numerical results obtained at the late stage of the simulation are given in Section~\ref{sec:03}. Comparison of the performance of the new (vorticity) method and the SDF method is in Section~\ref{sec:04}. Conclusion is in Section~\ref{sec:05}. Appendix~\ref{app:01} describes the transitional stage of the simulation when the plasma temporarily forms a ring. Appendix~\ref{app:02} explains how the simulation parameters are selected.


\section{\label{sec:02} Model description}

The code uses particle-in-cell formalism in Cartesian coordinates. Electrons and ions are represented as charged particles. Multiple ion species can be introduced. The code resolves two spatial coordinates ($x$ and $y$) and three velocity components ($v_x$, $v_y$, and $v_z$) for each charged particle. The simulation domain is a rectangle. For electrostatic simulations, the domain can be completely enclosed, periodic in one direction only, or periodic in two directions; rectangular metal or dielectric objects can be introduced inside the domain. For electromagnetic simulation, the domain is completely enclosed, without metal objects inside. Including metal objects inside the domain in electromagnetic simulation requires additional steps which will be described in a separate publication.

The code includes models of electron-neutral collisions, charge-exchange ion-neutral collisions,\cite{MaiorovPPR2009} and electron emission from material surfaces due to electron and/or ion bombardment, similar to the models used in other codes developed by the authors.\cite{SydorenkoThesis,edipic-1d-web-page,edipic-2d-web-page}

The code is written in Fortran and parallelized with MPI. Domain decomposition is used. Linear equation systems defining components of the electrostatic and electromagnetic field are solved using the PETSc library. To ensure efficient operation of parallel field solvers, the whole domain is divided into sub-domains (referred to as blocks) of the same size. The number of blocks equals the number of MPI processes. The code includes an algorithm ensuring approximately even numerical load of each MPI process at the particle processing stage. Particles are advanced within sub-domains (referred to as clusters) combining several adjacent blocks. Particles belonging to a single cluster are processed by several MPI processes. The number of these processes changes dynamically, depending on the number of particles in the cluster, and is always no less than one.

The code uses four staggered structured uniform grids, similar to the Yee grids for electromagnetic simulation.\cite{YeeIEEETAP1966} The first grid has nodes with coordinates $x_i=i\Delta x$, $y_j=j\Delta x$, where $i$ and $j$ are integer numbers, $\Delta x$ is the cell size. Below, this grid is referred to as the centered grid. The second grid is shifted in the $x$-direction by half-a-cell relative to the centered grid, it has nodes with coordinates $x_{i+1/2}=(i+1/2)\Delta x$, $y_j=j\Delta x$. The third grid is shifted in the $y$-direction and has nodes with coordinates $x_i=i\Delta x$, $y_{j+1/2}=(j+1/2)\Delta x$. The fourth grid is shifted in both $x$ and $y$ directions, coordinates of its nodes are $x_{i+1/2}=(i+1/2)\Delta x$, $y_{j+1/2}=(j+1/2)\Delta x$.
In this paper, a node of a grid is specified by the pair of factors before the $\Delta x$ in the node coordinates $x$ and $y$ introduced above. One or both of theses factors may be half-integer for a node of one of the shifted grids. For instance, node $(i,j)$ belongs to the centered grid and has coordinates $x_i$ and $y_j$, node $(i+1/2,j)$ belongs to the second grid and has coordinates $x_{i+1/2}$ and $y_j$, etc.
Boundaries of the simulation domain and material objects inside the domain are aligned with the nodes of the centered grid. The electrostatic potential $\Phi$, charge density $\rho$, vector components of electric field $E_z$ and electric current density $J_z$ are defined on the centered grid. Vector components of electric field $E_x$, magnetic field $B_y$, and electric current density $J_x$ are defined on the second grid. Vector components of electric field $E_y$, magnetic field $B_x$, and electric current density $J_y$ are defined on the third grid. Vector component of magnetic field $B_z$ is defined on the fourth grid.

Below, the electric field and current are often separated into their solenoidal ($\vec{E}_{sol}$, $\vec{J}_{sol}$) and irrotational ($\vec{E}_{irr}$, $\vec{J}_{irr}$) parts. This means that $\vec{E}=\vec{E}_{sol}+\vec{E}_{irr}$ where $\nabla\cdot\vec{E}_{sol}=0$ and $\nabla\times\vec{E}_{irr}=0$. Similarly, $\vec{J}=\vec{J}_{sol}+\vec{J}_{irr}$ where $\nabla\cdot\vec{J}_{sol}=0$ and $\nabla\times\vec{J}_{irr}=0$.


\subsection{\label{sec:02-1} Electrostatic part}

The direct implicit algorithm is well described in Ref.~\onlinecite{GibbonsJCP1995}. A brief description of this algorithm is provided below for convenience.

The numerical scheme is of a "leap-frog" type, with particle coordinates and velocities defined at times shifted by half of a time step $\Delta t$. In this paper, superscript $n$ denotes values defined at time $t^n=n\Delta t$, superscript $n\pm 1/2$ denotes values defined at time $t^{n\pm 1/2}=t^n\pm\Delta t/2$, subscript $p$ denotes values related to a single particle. In the beginning of a computational cycle advancing the system from time $t^n$ to time $t^{n+1}$, one knows grid values of solenoidal electric field $\vec{E}^n_{sol}$, irrotational electric field $\vec{E}_{irr}^n$, magnetic field $\vec{B}^n$, particle coordinates $\vec{x}_p^n$, velocities $\vec{v}_p^{n-1/2}$, and accelerations $\vec{a}_p^{n-1}$. Particle advance is performed in two stages. First, particles are accelerated to "streaming" velocities and pushed to "streaming" positions:
%
%
\begin{equation}\label{eq:01}
    \tilde{\vec{v}}_p=
    \mathbf{R}^n\cdot\vec{v}_p^{n-1/2}
    +
    \dfrac{\Delta t}{2}
    \left(\mathbf{1}+\mathbf{R}^n\right)\cdot
    \left(\dfrac{1}{2}\vec{a}_p^{n-1}+
          \dfrac{q_s}{m_s}\vec{E}^n_{sol}\right)~,
\end{equation}
%
%
\begin{equation}\label{eq:02}
    \tilde{\vec{x}}_p=
    \vec{x}_p^n+\Delta t\tilde{\vec{v}}_p~,
\end{equation}
where $q_s$ and $m_s$ are the charge and mass of a particle, subscript $s$ denotes particle species, $\mathbf{1}$ is a unitary matrix, matrix $\mathbf{R}^n$ describes rotation in the magnetic field $\vec{B}^n$ and has coefficients
%
%
\begin{equation}\label{eq:03}
\begin{split}
R_{11}=\left(1-\theta^2+2\alpha^2 B_x^2\right)/(1+\theta^2)
\\
R_{12}=2\left(\alpha^2 B_x B_y+\alpha B_z\right)/(1+\theta^2)
\\
R_{13}=2\left(\alpha^2 B_x B_z-\alpha B_y\right)/(1+\theta^2)
\\
R_{21}=2\left(\alpha^2 B_x B_y-\alpha B_z\right)/(1+\theta^2)
\\
R_{22}=\left(1-\theta^2+2\alpha^2 B_y^2\right)/(1+\theta^2)
\\
R_{23}=2\left(\alpha^2 B_y B_z+\alpha B_x\right)/(1+\theta^2)
\\
R_{31}=2\left(\alpha^2 B_x B_z+\alpha B_y\right)/(1+\theta^2)
\\
R_{32}=2\left(\alpha^2 B_y B_z-\alpha B_x\right)/(1+\theta^2)
\\
R_{33}=\left(1-\theta^2+2\alpha^2 B_z^2\right)/(1+\theta^2)
\end{split}
\end{equation}
with $\alpha=q\Delta t/2m$ and $\theta^2=\alpha^2(B_x^2+B_y^2+B_z^2)$, here superscripts denoting time in $B_{x,y,z}$ and species subscripts in $q$ and $m$ are omitted for brevity.

Next, the "streaming" charge density $\tilde{\rho}$ is calculated in the nodes of the centered grid using "streaming" particle coordinates. The following modified Poisson equation is solved to find the advanced electrostatic potential $\Phi^{n+1}$:
%
%
\begin{equation}\label{eq:04}
    \nabla\cdot\left[\left(
    \varepsilon_r \mathbf{1}+\mathbf{X}
    \right)\cdot\nabla\Phi^{n+1}\right]
    =-\dfrac{\tilde{\rho}}{\varepsilon_0}~,
\end{equation}
where $\varepsilon_r$ is the relative dielectric constant, $\varepsilon_0$ is the electric constant, and $\mathbf{X}$ is the implicit susceptibility tensor defined as
%
%
\begin{equation}\label{eq:05}
    \mathbf{X}=\dfrac{\Delta t^2}{4}
    \sum_s\left(\mathbf{1}+\mathbf{R}^n\right)
    \dfrac{q_s\tilde{\rho}_s}{m_s \varepsilon_0}~,
\end{equation}
where the summation is performed over all charged species. Note that in plasma-filled areas $\varepsilon_r=1$ while inside the dielectric areas $\tilde\rho=0$ and $\mathbf{X}=0$.

In the two-dimensional (2D) system, only four components of tensor $\mathbf{X}$ are required: $X_{11}$, $X_{12}$, $X_{21}$, and $X_{22}$. Like the charge density, these components are defined in nodes of the centered grid. Finite difference representation of Eq.~\ref{eq:04}, however, requires $X_{11}$ and $X_{12}$ in nodes of the grid shifted in the $x$-direction and $X_{21}$ and $X_{22}$ in nodes of the grid shifted in the $y$-direction. The code calculates the aforementioned tensor components in the centered grid nodes, tensor values in the shifted nodes are obtained by averaging of values in the two neighbor centered nodes. The electrostatic field is obtained as $\vec{E}_{irr}^{n+1}=-\nabla\Phi^{n+1}$. In finite differences, formulas for the electric field are
%
%
\begin{equation}\label{eq:06}
\begin{split}
    E_{irr,x;i+1/2,j}^{n+1}=(\Phi_{i,j}^{n+1}-\Phi_{i+1,j}^{n+1})/\Delta x, \\
    E_{irr,y;i,j+1/2}^{n+1}=(\Phi_{i,j}^{n+1}-\Phi_{i,j+1}^{n+1})/\Delta x~.
\end{split}
\end{equation}

The modified Poisson's equation (\ref{eq:04}) is solved with the values of the electrostatic potential given at metal walls enclosing the simulation domain (below referred to as the external walls) and at the surfaces of metal objects placed inside the simulation domain (below referred to as the inner objects). It is assumed that the metal is a perfect conductor, the potential is constant along the surface of metal walls and inner objects. Metal walls with different potential may be connected by segments with a linear potential profile along their surface, with the end values corresponding to the potentials of the metal walls. In electrostatic simulation, periodic boundary conditions can be applied at the opposite sides of the simulation domain.

At the second stage of particle advance, the electrostatic field is used to finish acceleration of particles to their final velocities:
%
%
\begin{equation}\label{eq:07}
    \delta\vec{v}_p=
    \dfrac{q_s\Delta t}{4 m_s}
    \left(\mathbf{1}+\mathbf{R}^n\right)
    \vec{E}_{irr}^{n+1}~,
\end{equation}
%
%
\begin{equation}\label{eq:08}
    \vec{v}_p^{n+1/2}=\tilde{\vec{v}}_p+\delta\vec{v}_p~.
\end{equation}
In (\ref{eq:07}), the electrostatic field is taken in the "streaming" position $\tilde{\vec{x}}_p$ of the particle. Then the particles are pushed to their final positions:
%
%
\begin{equation}\label{eq:09}
    \vec{x}_p^{n+1}=\tilde{\vec{x}}_p+\Delta t\delta\vec{v}_p~.
\end{equation}
After this, the particle acceleration is updated as follows:
%
%
\begin{equation}\label{eq:10}
    \vec{a}_p^{n}=\dfrac{1}{2}
    \left(\vec{a}_p^{n-1}+\dfrac{q_s}{m_s}\vec{E}_{irr}^{n+1}\right)~,
\end{equation}
where the electrostatic field is taken in the particle's final position $\vec{x}_p^{n+1}$. This concludes the electrostatic part of the algorithm.


\subsection{\label{sec:02-2} Electromagnetic part}

The Darwin method uses the following equations for solenoidal electric and self-consistent magnetic fields:
%
%
\begin{equation}\label{eq:11}
    \nabla\times\vec{B}=\mu_0\vec{J}_{sol}~,
\end{equation}
%
%
\begin{equation}\label{eq:12}
    \nabla^2\vec{E}_{sol}=\mu_0\dfrac{\partial\vec{J}_{sol}}{\partial t}~,
\end{equation}
where $\mu_0$ is the magnetic constant and $\vec{J}_{sol}$ is the solenoidal part of the electric current. Equation (\ref{eq:11}) follows from Ampere-Maxwell law where the solenoidal part of the displacement current ${c^{-2}}{\partial\vec{E}_{sol}}/{\partial t}$ is omitted and relationship
%
%
\begin{equation}\label{eq:13}
    \dfrac{\partial\vec{E}_{irr}}{\partial t}=
    -\dfrac{\vec{J}_{irr}}{\varepsilon_0}
\end{equation}
is applied. Equation (\ref{eq:13}) follows from Gauss's law and charge continuity. Equation (\ref{eq:12}) is obtained by applying curl operator to Faraday's law and using (\ref{eq:11}).

Below, various moments of the velocity distribution function are calculated, which requires to have particle velocities defined at the same time as the final coordinates (\ref{eq:09}). These velocities are obtained after the end of the electrostatic part as follows:
%
%
\begin{equation}\label{eq:14}
    \vec{v}_p^{n+1}=
    \mathbf{R}^n\cdot\vec{v}_p^{n+1/2}
    +
    \dfrac{q_s\Delta t}{4 m_s}
    \left(\mathbf{1}+\mathbf{R}^n\right)\cdot
    \left(\vec{E}_{irr}^{n+1}+\vec{E}_{sol}^n\right)~.
\end{equation}
Note that velocities (\ref{eq:14}) are not used to advance particle coordinates. The full predicted electric current $\vec{J}_{pred}$ is calculated as
%
%
\begin{equation}\label{eq:15}
    \vec{J}_{pred,g}=\sum_s{q_s\sum_p{S(\vec{x}_g-\vec{x}_p^{n+1})\vec{v}_p^{n+1}}}~,
\end{equation}
where subscript $g$ denotes grid values, summation is over all particles (subscript $p$) of all charged species (subscript $s$), $\vec{x}_g$ is the grid node coordinate vector, $S$ is the function defining particle contribution to the density in surrounding nodes (corresponds to the bilinear interpolation in 2D). The time derivative of the total electric current is calculated as
%
%
\begin{equation}\label{eq:16}
    \mu_0\frac{\partial\vec{J}}{\partial t}
    =
    \mu\vec{E}_{sol} + \vec{Q}~,
\end{equation}
where
%
%
\begin{equation}\label{eq:17}
\begin{split}
    \mu&=\sum_s{\dfrac{\omega_{pl,s}^2}{c^2}}~,
    \\
    \vec{Q}&=\mu\vec{E}_{irr}+\vec{\zeta}\times\vec{B}+\vec{K}+\vec{C}~,
    \\
    \vec{\zeta}&=\sum_s{\dfrac{\omega_{pl,s}^2}{c^2}\vec{u}_s}~,
    \\
    \vec{K}&=-\mu_0\sum_s{q_s\nabla\left(n_s\langle\vec{v}_s\vec{v}_s\rangle\right)}~,
    \\
    \vec{C}&=-\mu_0\sum_s{q_s n_s \vec{u}_s \nu_s}~,
\end{split}
\end{equation}
and $\omega_{pl,s}^2$ is the plasma frequency, $\vec{v}_s$ is the random velocity, $\vec{u}_s$ = $\langle\vec{v}_s\rangle$ is the flow velocity, $n_s$ is the number density, and $\nu_s$ is the momentum transfer collision frequency of charged species $s$, the summation is performed over all charged species. Field solver for $\vec{E}_{sol}$ requires grid quantities $\mu_g$ and $\vec{Q}_g$ which are obtained using direct accumulation of particle quantities in grid nodes:
%
%
\begin{equation}\label{eq:18}
\begin{split}
    \mu_g &=\mu_0\sum_s{\dfrac{q_s^2}{m_s}\sum_p{S(\vec{x}_g-\vec{x}_p^{n+1})}}~,
    \\
    \vec{Q}_g&=(\mu\vec{E}_{irr}^{n+1})_g+(\vec{\zeta}\times\vec{B}^n)_g+\vec{K}_g+\vec{C}_g~,
    \\
    (\mu\vec{E}_{irr}^{n+1})_g &=
    \mu_0\sum_s{\dfrac{q_s^2}{m_s}\sum_p
    {S(\vec{x}_g-\vec{x}_p^{n+1})\vec{E}_{irr}^{n+1}(\vec{x}_p^{n+1})}}~,
    \\
    (\vec{\zeta}\times\vec{B}^{n})_g &=\mu_0\sum_s
    {\dfrac{q_s^2}{m_s}\sum_p
    {S(\vec{x}_g-\vec{x}_p^{n+1})\vec{v}_p^{n+1}\times\vec{B}^{n}(\vec{x}_p^{n+1})}}~,
    \\
    \vec{K}_g &=-\mu_0\nabla\cdot
    \sum_s{q_s\sum_p{\vec{v}_p^{n+1}\vec{v}_p^{n+1} S(\vec{x}_g-\vec{x}_p^{n+1})}}~,
    \\
    \vec{C}_g &=\mu_0\Delta t^{-1}\sum_s{q_s\sum_{p_c}{\delta\vec{v}_{p_c}S(\vec{x}_g-\vec{x}_{p_c}^{n+1})}}~.
\end{split}
\end{equation}
Term $\vec{C}_g$ in (\ref{eq:18}) represents variation of the electric current due to collisions. In this term, summation over $p_c$ is summation over particles collided at this time step, $\delta\vec{v}_{p_c}=\vec{v}_{p_c}^a-\vec{v}_{p_c}^b$ is the particle velocity change in the collision, $\vec{v}_{p_c}^b$ and $\vec{v}_{p_c}^a$ are the particle velocity vectors before and after the collision. Collision procedures for electrons and ions are involved after the electrostatic part of the algorithm is completed. Contributions to $\vec{C}_g$ are accounted for every time a collision occurs.

It is instructive to mention that Equation (\ref{eq:16}) follows from the first velocity moment of Vlasov equation
%
%
\begin{equation}\label{eq:19}
    \dfrac{\partial\left(n m \vec{u}\right)}{\partial t}
    =
    n q \left( \vec{E} + \vec{u}\times\vec{B} \right) -
    \nabla\left(m n \langle\vec{v}\vec{v}\rangle\right) - m n \vec{u} \nu~.
\end{equation}
This approach is necessary because, as mentioned in Ref.~\onlinecite{GibbonsJCP1995}, finite time differencing cannot be applied in (\ref{eq:12}) because it causes strong numerical instability.

Equation (\ref{eq:11}) requires the solenoidal part of the current. Since any vector normal to the simulation plane is solenoidal, one can use the $z$-component of current (\ref{eq:15}) directly in the $z$-component of (\ref{eq:11}) specifying $B_x$ and $B_y$. Situation with the $B_z$ is less straightforward. The solenoidal current in the simulation plane required to find $B_z$ can be obtained from $\vec{J}_{sol}=\vec{J}_{pred}-\nabla\phi$ where scalar function $\phi$ satisfies $\nabla^2\phi=\nabla\cdot\vec{J}_{pred}$. The corresponding equation
%
%
\begin{equation}\label{eq:20}
    \left(\dfrac{\partial^2}{\partial x^2}+
          \dfrac{\partial^2}{\partial y^2}\right) \phi
    =
    \dfrac{\partial J_{pred,x}}{\partial x}+
    \dfrac{\partial J_{pred,y}}{\partial y}
\end{equation}
is solved with boundary condition $\phi=0$ at the domain boundary.\cite{StartsevPAC2007} The solenoidal current components are found as
%
%
\begin{equation}\label{eq:21}
\begin{split}
    J_{sol,x}&=J_{pred,x}-\dfrac{\partial\phi}{\partial x}~, \\
    J_{sol,y}&=J_{pred,y}-\dfrac{\partial\phi}{\partial y}~.
\end{split}
\end{equation}

In Ref.~\onlinecite{GibbonsJCP1995}, the magnetic field normal to the simulation plane is calculated by taking a curl of Equation (\ref{eq:11}) which eliminates the need to calculate the solenoidal electric current. However, it is difficult to formulate a practical boundary condition for $B_z$ on metal surfaces since the surface electric current is in general not known. Therefore, to calculate the magnetic field, a vector potential $\vec{A}$ is introduced such that
%
%
\begin{equation}\label{eq:22}
    \vec{B}=\nabla\times\vec{A}~,
\end{equation}
and $\vec{A}$ satisfies the Coulomb gauge, $\nabla\cdot\vec{A}=0$. With (\ref{eq:22}), equation (\ref{eq:11}) gives
%
%
\begin{equation}\label{eq:23}
    \left(\dfrac{\partial^2}{\partial x^2}+
          \dfrac{\partial^2}{\partial y^2}\right) A_x=-\mu_0 J_{sol,x}~,
\end{equation}
%
%
\begin{equation}\label{eq:24}
    \left(\dfrac{\partial^2}{\partial x^2}+
          \dfrac{\partial^2}{\partial y^2}\right) A_y=-\mu_0 J_{sol,y}~,
\end{equation}
and
%
%
\begin{equation}\label{eq:25}
    \left(\dfrac{\partial^2}{\partial x^2}+
          \dfrac{\partial^2}{\partial y^2}\right) A_z=-\mu_0 J_{pred,z}~.
\end{equation}

Equation (\ref{eq:23}) is solved with boundary conditions $A_x=0$ at $y=0$ and $y=L_y$ and $\partial A_x/\partial x=0$ at $x=0$ and $x=L_x$, where $L_x$ and $L_y$ are the domain length along the $x$ and $y$-directions, respectively. Equation (\ref{eq:24}) is solved with boundary conditions $A_y=0$ at $x=0$ and $x=L_x$ and $\partial A_y/\partial y=0$ at $y=0$ and $y=L_y$. These boundary conditions ensure that circulation of the vector potential along the boundary of the simulation domain is zero, which means that the flux of $B_z$ through the domain area is zero. The latter implies that the circulation of the electric field along the domain boundary is also zero which matches the boundary condition for $E_{sol;x,y}$ introduced below. Equation (\ref{eq:25}) is solved with boundary conditions $A_z=0$ at the entire domain boundary. This condition ensures that the magnetic field in the simulation plane is normal to the surface of the domain.

The magnetic field components are calculated as
%
%
\begin{equation}\label{eq:26}
\begin{split}
    B_x^{n+1}&=\dfrac{\partial A_z}{\partial y}~, \\
    B_y^{n+1}&=-\dfrac{\partial A_z}{\partial x}~, \\
    B_z^{n+1}&=\dfrac{\partial A_y}{\partial x}-\dfrac{\partial A_x}{\partial y}~.
\end{split}
\end{equation}
It is necessary to mention that the values of $A_x$ are defined in the nodes of the grid shifted along the $x$-direction (same as $J_x$), $A_y$ are defined in the nodes of the grid shifted along the $y$-direction (same as $J_y$), and $A_z$ are defined in the nodes of the centered grid (same as $J_z$).

The $x$ and $y$ components of (\ref{eq:16}) contain both the solenoidal and the irrotational parts. Separating these parts without knowing $E_{sol,x}$ and $E_{sol,y}$ is not straightforward. The SDF  method applied in Ref.~\onlinecite{GibbonsJCP1995} assumes that
%
%
\begin{equation}\label{eq:27}
    \mu_0\dfrac{\partial\vec{J}_{irr}}{\partial t}=-\nabla^2\nabla\psi~,
\end{equation}
where $\psi$ is some scalar function, introduces
%
%
\begin{equation}\label{eq:28}
    \vec{\Xi}=\vec{E}_{sol}-\nabla\psi~,
\end{equation}
solves a coupled system of partial differential equations
%
%
\begin{equation}\label{eq:29}
\begin{split}
    \nabla^2\vec{\Xi}-\mu\vec{\Xi}&=\vec{Q}+\mu\nabla\psi~,
    \\
    \nabla^2\psi&=-\nabla\cdot\vec{\Xi}~,
\end{split}
\end{equation}
and finally calculates the solenoidal electric field as
%
%
\begin{equation}\label{eq:30}
    \vec{E}_{sol}^{n+1}=\vec{\Xi}+\nabla\psi~.
\end{equation}
In the selected 2D geometry, all vectors and vector differential operators in (\ref{eq:27}-\ref{eq:30}) have only the $x$ and $y$ components.

The SDF method is implemented in the code described in this paper as one of two options for calculation of $E_{sol,x}^{n+1}$ and $E_{sol,y}^{n+1}$. The linear system of equations corresponding to the finite difference form of (\ref{eq:29}) is solved using the biconjugate gradient stabilized algorithm (KSPBCGSL solver of the PETSc library) with a multigrid preconditioner (PCMG method of the PETSc library). After trying various types of solvers and preconditioners available in PETSc, it was found that this combination gives the best results. However, solving this system does not always go smooth, as discussed in Section~\ref{sec:04}. In Refs.~\onlinecite{GibbonsJCP1995,HewettJCP1992}, such an equation system is solved using the dynamic alternating direction implicit (DADI) method. In the present paper, the following alternative approach is described.

One can apply the curl operator to (\ref{eq:12}) and use $\nabla\times\vec{J}_{sol}=\nabla\times\vec{J}$ to obtain
%
%
\begin{equation}\label{eq:31}
    \nabla^2\nabla\times\vec{E}_{sol}^{n+1}=
    \mu_0\nabla\times\dfrac{\partial\vec{J}}{\partial t}~.
\end{equation}
In order to ensure that $\nabla\cdot\vec{E}_{sol}=0$, one can assume
%
%
\begin{equation}\label{eq:32}
    \vec{E}_{sol}^{n+1}=\nabla\times\vec{F}~,
\end{equation}
where $\vec{F}$ is some vector function. Substituting (\ref{eq:16}) and (\ref{eq:32}) into Equation (\ref{eq:31}) one obtains
%
%
\begin{equation}\label{eq:33}
    \nabla^2\left(\nabla\times\nabla\times\vec{F}\right)=
    \nabla\times\left(\mu\nabla\times\vec{F}+\vec{Q}\right)~.
\end{equation}
In the selected 2D geometry, it is sufficient to know only the $z$ component of $\vec{F}$. Below, for the sake of brevity, this component is denoted $F$, with subscript $z$ omitted.
Then (\ref{eq:33}) becomes
%
%
\begin{equation}\label{eq:34}
\begin{split}
    \left(\dfrac{\partial^2}{\partial x^2}+
          \dfrac{\partial^2}{\partial y^2}\right)
    \left(\dfrac{\partial^2}{\partial x^2}+
          \dfrac{\partial^2}{\partial y^2}\right) F
           -
           \\
    \left(\dfrac{\partial}{\partial x}\mu\dfrac{\partial}{\partial x}+
          \dfrac{\partial}{\partial y}\mu\dfrac{\partial}{\partial y}\right) F
           =
           \dfrac{\partial Q_x}{\partial y}-
           \dfrac{\partial Q_y}{\partial x}~.
\end{split}
\end{equation}
Once the $F$ values are known, the solenoidal electric field components in the simulation plane are found as
%
%
\begin{equation}\label{eq:35}
\begin{split}
    {E}_{sol,x}^{n+1}&=\dfrac{\partial F}{\partial y}~,
    \\
    {E}_{sol,y}^{n+1}&=-\dfrac{\partial F}{\partial x}~.
\end{split}
\end{equation}

\begin{figure}
\includegraphics{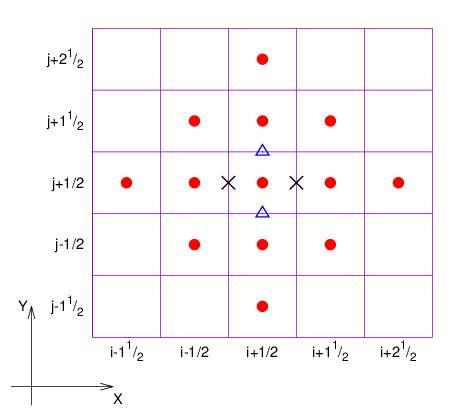}
\caption{\label{fig:01}
Markers show positions of nodes of numerical grids on the coordinate plane $x-y$. The solid red circles belong to the grid where $F$ is defined. Nodes of the grid for $Q_x$ and $Q_y$ used in the right-hand side of (\ref{eq:36}) are shown by the blue triangles and black diagonal crosses, respectively. Note that the four coefficients $\mu$ in the left hand side of (\ref{eq:36}) are defined in the same nodes as the $Q_x$ and $Q_y$. The solid purple lines go through nodes of the centered grid.
}
\end{figure}

Equation (\ref{eq:34}) creates a system of linear equations for values of $F$ defined in the nodes of the grid shifted relative to the centered grid in both the $x$ and $y$ directions. Each equation of this system involves 13 nodes of the $F$ grid positioned according to the stencil shown by red circles in Figure~\ref{fig:01}. In finite difference form, Equation (\ref{eq:34}) with the central node of the stencil in node $(i+1/2,j+1/2)$ of the $F$ grid is
%
%
\begin{equation}\label{eq:36}
\begin{split}
    F_{i+\tfrac{1}{2},j+2\tfrac{1}{2}}+
    \\
    2 F_{i-\tfrac{1}{2},j+1\tfrac{1}{2}}+
    \\
    \left(-8-\Delta x^2\mu_{i+\tfrac{1}{2},j+1}\right) F_{i+\tfrac{1}{2},j+1\tfrac{1}{2}}+
    \\
    2 F_{i+1\tfrac{1}{2},j+1\tfrac{1}{2}}+
    \\
    F_{i-1\tfrac{1}{2},j+\tfrac{1}{2}}+
    \\
    \left(-8-\Delta x^2\mu_{i,j+1/2}\right) F_{i-\tfrac{1}{2},j+\tfrac{1}{2}}+
    \\
    \left[20+\Delta x^2\left(\mu_{i+\tfrac{1}{2},j+1}\right.\right. + &
    \\
    \left.\left.\mu_{i,j+\tfrac{1}{2}}+\mu_{i+1,j+\tfrac{1}{2}}+\mu_{i+\tfrac{1}{2},j}\right) \right] F_{i+\tfrac{1}{2},j+\tfrac{1}{2}}+
    \\
    \left(-8-\Delta x^2\mu_{i+1,j+\tfrac{1}{2}}\right) F_{i+1\tfrac{1}{2},j+\tfrac{1}{2}}+
    \\
    F_{i+2\tfrac{1}{2},j+\tfrac{1}{2}}+
    \\
    2 F_{i-\tfrac{1}{2},j-\tfrac{1}{2}}+
    \\
    \left(-8-\Delta x^2\mu_{i+\tfrac{1}{2},j}\right) F_{i+\tfrac{1}{2},j-\tfrac{1}{2}}+
    \\
    2 F_{i+1\tfrac{1}{2},j-\tfrac{1}{2}}+
    \\
    F_{i+\tfrac{1}{2},j-1\tfrac{1}{2}}=
    \\
    \Delta x^3\left( Q_{x;i+\tfrac{1}{2},j+1} - Q_{x;i+\tfrac{1}{2},j}+
                     Q_{y;i,j+\tfrac{1}{2}} - Q_{y;i+1,j+\tfrac{1}{2}} \right)
\end{split}
\end{equation}

Note that since the $F$ grid nodes are shifted along both directions, they are not on the domain boundary aligned with the nodes of the centered grid. The field solver includes $F$ grid  nodes which are half-a-cell outside the domain. The boundary conditions for (\ref{eq:36}) require that $F=0$ in the nodes half-a-cell outside the domain and in the nodes which are inside the domain and half-a-cell from the domain boundary. With (\ref{eq:35}), this condition ensures that both $E_{sol,x}$ and $E_{sol,y}$ are zero at the domain boundary, similar to Ref.\onlinecite{GibbonsJCP1995}.
The system of equations (\ref{eq:36}) is solved using the KSPBCGSL solver of the PETSc library with the multigrid preconditioner from the LLNL package HYPRE. Since the left-hand side of (\ref{eq:31}) contains the curl of $\vec{E}_{sol}$, below the method of calculation of $E_{sol;x,y}$ involving Equations (\ref{eq:34}-\ref{eq:36}) is referred to as the vorticity method.
%
%

The $z$-component of the time derivative of the predicted electric current (\ref{eq:16}) is solenoidal. In combination with (\ref{eq:12}), it gives the following equation for the solenoidal electric field normal to the simulation plane:
%
%
\begin{equation}\label{eq:37}
     \left(\dfrac{\partial^2}{\partial x^2}+
      \dfrac{\partial^2}{\partial y^2} - \mu\right)E_{sol,z}^{n+1}=Q_z~.
\end{equation}
The boundary condition for (\ref{eq:37}) is $E_{sol,z}=0$ at the domain surface. Systems of linear equations corresponding to Equations (\ref{eq:04}), (\ref{eq:20}), (\ref{eq:23}-\ref{eq:25}), and (\ref{eq:37}) are solved using the Generalized Minimal Residual Method (KSPGMRES solver of the PETSc library) with the HYPRE multigrid preconditioner.

In an ICP simulation, the simulation domain is completely enclosed by grounded metal walls. An antenna driving the ICP is represented by a region where the electric current density is a given function of time, $\vec{J}^{ext}(t)$. This external current must have zero divergence. Such a region is considered transparent to the electromagnetic field. Below, it is referred to as the current patch. The electric current in the patch can be in the simulation plane, i.e. has components $J^{ext}_x$ and $J^{ext}_y$, or be normal to the simulation plane, i.e. has component $J^{ext}_z$ only. With the external currents, the right hand side (RHS) of Equations (\ref{eq:23}-\ref{eq:25}) acquire an additional term $-\mu_0 J^{ext}_{x,y,z}$, respectively. The RHS of Equation (\ref{eq:34}) acquires an additional term
%
%
\begin{equation}\label{eq:38}
\mu_0\left(\dfrac{\partial}{\partial y}\dfrac{\partial J^{ext}_x}{\partial t}
          -\dfrac{\partial}{\partial x}\dfrac{\partial J^{ext}_y}{\partial t}\right)~.
\end{equation}
The RHS of Equation (\ref{eq:37}) acquires an additional term $\mu_0 \dfrac{\partial}{\partial t}J_z^{ext}$. Note that $\partial \vec{J}^{ext}/\partial t$ is also a known function of time.


\section{\label{sec:03} Simulation of an inductively coupled plasma}

\subsection{\label{sec:03-1} System setup and numerical parameters}

%
%
\begin{figure}
\includegraphics{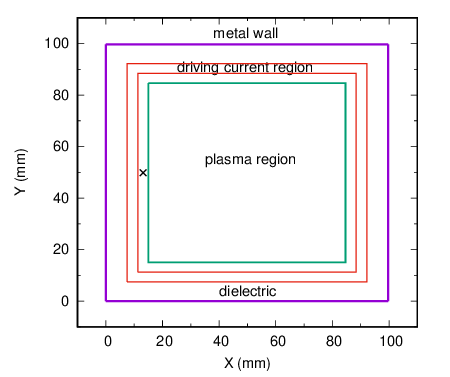}
\caption{\label{fig:02}
Setup for simulation of an inductively coupled plasma with the driving current in the $x$-$y$ plane. The cross marks the probe where the dependencies of amplitude and phase shift of $E_{sol,y}$ versus time shown in Figure~\ref{fig:03}(b) are obtained.
}
\end{figure}
The simulation domain is a square surrounded by metal walls (purple square in Figure~\ref{fig:02}). A relatively thin layer of dielectric with dielectric constant $\varepsilon_r=1$ is placed along the walls leaving an empty square area in the center of the system (shown by the green line in Figure~\ref{fig:02}), this is the area where the plasma and the neutral gas exist. Inside the dielectric layer, there is a region with the driving electric current density $J_{x,y}^{ext}$ bounded between two squares (shown by red lines in Figure~\ref{fig:02}). The driving electric current flows in a closed loop around the plasma region. The electric current density components are defined as
%
%
\begin{equation}\label{eq:39}
\begin{split}
    J_x^{ext}(x,y,t)=I_1\sin(\omega t)\dfrac{\partial}{\partial y}g(x,y)~, \\
    J_y^{ext}(x,y,t)=-I_1\sin(\omega t)\dfrac{\partial}{\partial x}g(x,y)~,
\end{split}
\end{equation}
where $I_1$ is the amplitude of the total electric current flowing through the region's cross-section per 1 meter of length along the $z$-direction (in units of A/m), $\omega$ is the driving current frequency, and function $g$ is zero outside the external boundary of the current region, $g=1$ inside the internal boundary of the current region, and inside the current region it is a solution of Poisson's equation
%
%
\begin{equation}\label{eq:40}
    \left(\dfrac{\partial^2}{\partial x^2}+
          \dfrac{\partial^2}{\partial y^2}\right) g=0~,
\end{equation}
solved with boundary conditions $g=0$ and $g=1$ at the external and internal boundaries of the driving current region, respectively. Note that the electric current with components defined by (\ref{eq:39}) has zero divergence.

The cell size of the numerical grid $\Delta x$ and the time step $\Delta t$ are calculated from the following input parameters: the scale electron density $n_{e,0}$, the scale electron temperature $T_{e,0}$ (in units of eV), the ratio of the scale Debye length and the grid cell size $\alpha$, and the ratio of the maximal allowed particle velocity and the scale electron thermal velocity $\beta$. The scale Debye length is calculated as $\lambda_{e,0}={v_{th,e,0}}/{\omega_{e,0}}$ where $v_{th,e,0}=\sqrt{{2eT_{e,0}}/{m_e}}$ is the scale electron thermal velocity, and
$\omega_{e,0}=\sqrt{{n_{e,0}e^2}/{m_e\varepsilon_0}}$ is the scale electron plasma frequency. The grid cell size and the time step are calculated as
%
%
\begin{equation}\label{eq:41}
    \Delta x=\dfrac{\lambda_{e,0}}{\alpha},~\Delta t=\dfrac{\Delta x}{\beta v_{th,e,0}}~.
\end{equation}

In the simulation described in this Section, $n_{e,0}=1.6\times 10^{18}\text{~m}^{-3}$, $T_{e,0}=4\text{~eV}$, $\alpha=0.044194$ (the exact value is $\sqrt{2}/32$), $\beta=9$, $\omega_{e,0}=7.136\times 10^{10}\text{~s}^{-1}$, $v_{th,e,0}=1.186\times 10^6\text{~m/s}$, $\lambda_{e,0}=1.662\times 10^{-5}\text{~m}$, $\Delta x=3.761\times 10^{-4}\text{~m}$, and $\Delta t=3.523\times 10^{-11}\text{~s}$. The number of particles per cell for the scale density is $N_{ppc}=8000$. For a plasma with density and temperature as the scale values, the grid does not resolve the Debye length ($\Delta x/\lambda_{e,0}=22.63$) and the time step barely resolves the period of electron plasma oscillations ($\omega_{e,0}\Delta t=2.51$). The grid, however, resolves the skin layer length which for the plasma with the scale density is $c/\omega_{e,0}=4.2\times 10^{-3}\text{~m}$. The reasoning behind selection of $\beta$ and $N_{ppc}$ is given in Appendix~\ref{app:02}.

The centered grid has 266 nodes along the $x$ and $y$ directions. Correspondingly, the size of each side of the domain is $99.7\text{~mm}$, see Figure~\ref{fig:02}. The square plasma region in the center has $186\times 186$ nodes, the size of one side of the region is $69.6\text{~mm}$. The external and internal square boundaries of the driving current region are $226\times 226$ and $206\times 206$ nodes, respectively. The sides of the external and internal driving current region boundaries are $84.6\text{~mm}$ and $77.1\text{~mm}$, respectively. The amplitude of the driving current is $I_1=1000\text{~A/m}$. The frequency of the driving current is $10\text{~MHz}$. The plasma region is filled with uniform neutral gas Argon of density $10^{21}\text{~m}^{-3}$ and temperature $300\text{~K}$, the corresponding pressure is $31\text{~mTorr}$. Elastic, inelastic, and ionization collisions between electrons and neutral atoms are accounted for. Charge exchange collisions between ions and neutral atoms are accounted for. The ion mass is $40\text{~AMU}$. The simulation starts with the uniform plasma of density $10^{16}\text{~m}^{-3}$, electron temperature $4\text{~eV}$, ion temperature $0.03\text{~eV}$, initial electron and ion velocity distributions are Maxwellian. Simulation runs on a 96-core node of the Stellar cluster in PPPL.
%
%
\begin{figure}
\includegraphics{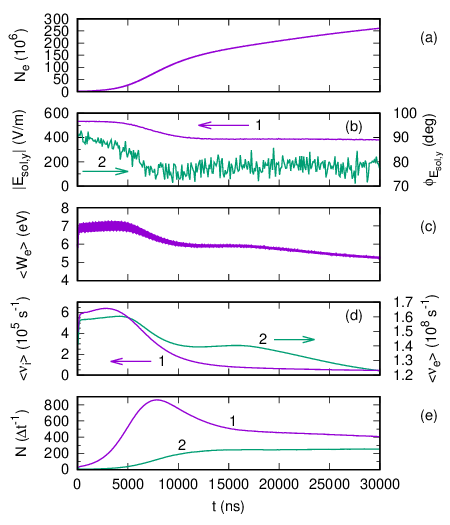}
\caption{\label{fig:03}
Time dependencies of (a) number of electron macroparticles, (b) amplitude (curve 1) and phase shift (curve 2) of $E_{sol,y}$ in probe shown by the black diagonal cross in Figure~\ref{fig:02}, (c) average kinetic electron energy, (d) ionization (curve 1) and elastic (curve 2) electron-neutral collisions frequencies, and (e) number of ionization collisions per time step (curve 1) and number of macroparticles escaped at the dielectric walls per time step (curve 2). Values in (d) and (e) are averaged over one driving current period ($100\text{~ns}$). In (b) and (d), curves 1 and 2 correspond to the left and right vertical coordinate axes, respectively (note the purple and green arrows pointing left and right). In (b), the phase shift (curve 2) is calculated relative to the phase of the driving electric current $I_1\sin(\omega t)$, i.e. it represents $E_{sol,y}\approx |E_{sol,y}|\sin(\omega t+\phi_{E_{sol,y}})$.
}
\end{figure}

\subsection{\label{sec:03-2} Results of simulation with the vorticity method}

In simulation described in this Section, the solenoidal electric field components $E_{sol;x,y}$ are calculated using the vorticity method. The simulation lasts for 30000 nanoseconds (ns). The plasma is sustained by the applied oscillating driving current. The number of macro-particles grows with time and by the end of simulation it exceeds 250 millions for each charged species, see Figure~\ref{fig:03}(a). The growth of the number of particles is initially exponential but it slows down after 10000 ns. While the plasma density rapidly grows between $t=5000\text{~ns}$ and $10000\text{~ns}$, the amplitude of the solenoidal electric field near the driving current region decreases (see curve 1 in Figure~\ref{fig:03}(b)) and the phase shift between the field and the driving electric current reduces from 90 degrees to about 80 degrees (see curve 2 in Figure~\ref{fig:03}(b)). In a real ICP device, such a change of the phase shift between the current and voltage in the antenna corresponds to the growth of irreversible power losses (Joule heating).

The average electron energy gradually decreases as more and more low-energy electrons are produced via ionization while the driving electromagnetic field penetrates only into a narrow boundary layer of the dense plasma, see Figure~\ref{fig:03}(c). The average frequency of ionization collisions rapidly decreases during the period of exponential growth of the number of particles, between 3000 ns and 10000 ns, see curve 1 in Figure~\ref{fig:03}(d). After that, the average ionization collision frequency decreases much slower, in the end of simulation it is about $4.6\times 10^4\text{~s}^{-1}$. The average frequency of elastic electron-neutral collisions is several orders of magnitude higher than the ionization frequency, compare curves 2 and 1 in Figure~\ref{fig:03}(d). Although the ionization collision frequency becomes almost constant, the system is not in the quasi-stationary state yet. Final particle losses at the walls are about 63\% of the total ionization rate, compare curves 1 and 2 in Figure~\ref{fig:03}(e). If the slopes of these curves remain the same as they are at $t=30000\text{~ns}$, the two curves will cross around $t=56000\text{~ns}$ and then the quasi-stationary state will be achieved. In this Section, the focus is on the plasma state at the late simulation stage ($t>27000~\text{~ns}$). Processes at the initial stage and transition between the two stages are discussed in Appendix~\ref{app:01}.
%
%
\begin{figure}
\includegraphics{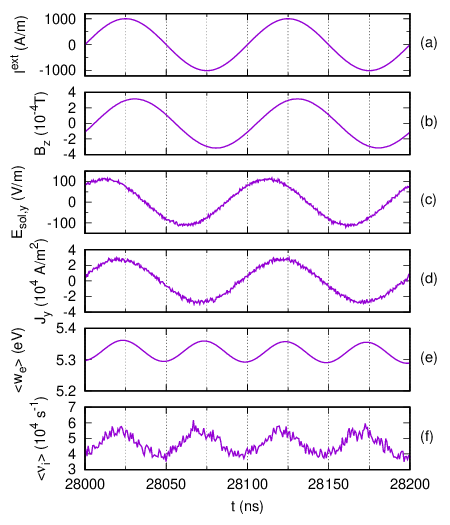}
\caption{\label{fig:04}
Driving electric current (a), magnetic field $B_z$ (b), solenoidal electric field $E_{sol,y}$ (c), plasma electric current density $J_y$ (d), average electron kinetic energy (e), and average ionization electron-neutral collision frequency (f) versus time. In (a), the positive values correspond to the current flowing in the counter-clockwise direction. Curves shown in (b,c,d) are obtained in a location (probe) shown by a diagonal cross in Figures~\ref{fig:05}(b), \ref{fig:06}(b), and \ref{fig:07}(b), respectively. In (e,f), the averaging is performed over all electron particles. Note that the collision frequency in (f) is not averaged over time like the frequencies in Figure~\ref{fig:03}(d).
}
\end{figure}

%
%
\begin{figure}
\includegraphics{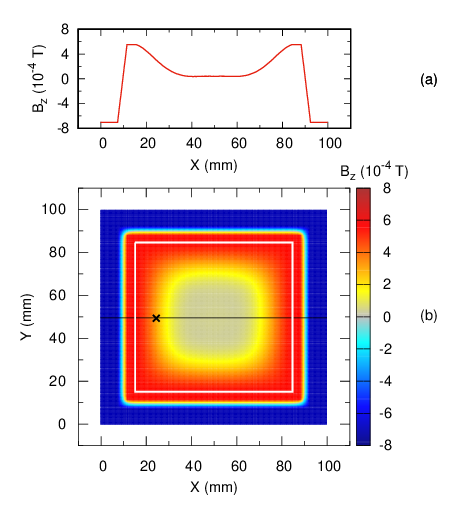}
\caption{\label{fig:05}
Magnetic field $B_z$ at time $t=28024.8\text{~ns}$. Panel (a) shows $B_z$ versus the $x$-coordinate in the cross-section with $y=50\text{~mm}$. Panel (b) shows the field versus the $x$ and $y$-coordinates. The cross-section shown in (a) is marked by the horizontal black solid line in (b). The diagonal cross marks position of a probe where the time dependence shown in Figure~\ref{fig:04}(b) is obtained. The white square marks boundaries of the plasma region.
}
\end{figure}

%
%
\begin{figure}
\includegraphics{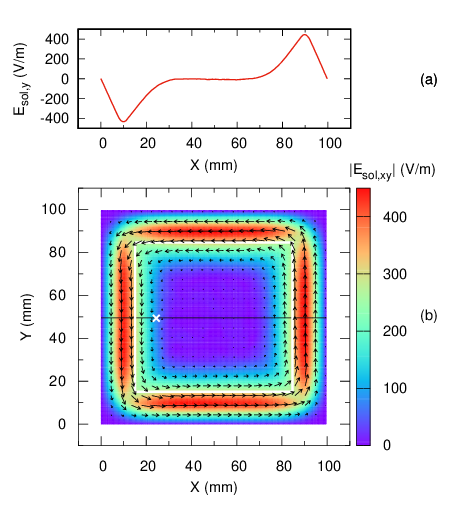}
\caption{\label{fig:06}
Solenoidal electric field at time $t=28049.6\text{~ns}$. Panel (a) shows $E_{sol,y}$ versus the $x$-coordinate in the cross-section with $y=50\text{~mm}$. Panel (b) shows the field versus the $x$ and $y$-coordinates. In (b), the colormap shows the absolute value of the field, $|E_{sol,xy}|=(E_{sol,x}^2+E_{sol,y}^2)^{1/2}$, the arrows show the direction of the vector. The cross-section shown in (a) is marked by the horizontal black solid line in (b). The diagonal cross marks position of a probe where the time dependence shown in Figure~\ref{fig:04}(c) is obtained. The white square marks boundaries of the plasma region.
}
\end{figure}

%
%
\begin{figure}
\includegraphics{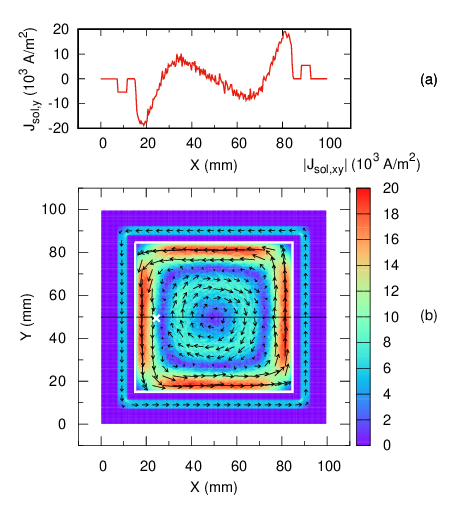}
\caption{\label{fig:07}
Solenoidal electric current density in the simulation plane at time $t=28049.6\text{~ns}$. Panel (a) shows $J_{sol,y}$ versus the $x$-coordinate in the cross-section with $y=50\text{~mm}$. Panel (b) shows the current versus the $x$ and $y$-coordinates. In (b), the color map shows the absolute value of the current, $|J_{sol,xy}|=(J_{sol,x}^2+J_{sol,y}^2)^{1/2}$, the arrows show the direction of the vector. The cross-section shown in (a) is marked by the horizontal black solid line in (b). The diagonal cross marks position of a probe where the time dependencies shown in Figure~\ref{fig:04}(d) is obtained. The white square marks boundaries of the plasma region.
}
\end{figure}

%
%
\begin{figure}
\includegraphics{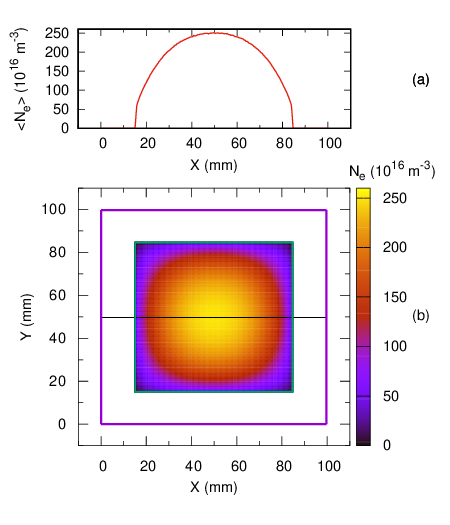}
\caption{\label{fig:08}
Profile of the electron density averaged over time interval $28000\text{~ns}$ - $28094\text{~ns}$ in a cross-section with $y=50\text{~mm}$.  (b) The density versus the $x$ and $y$-coordinates at time $t=28094.4\text{~ns}$. In (b), the purple and green squares mark the boundaries of the simulation domain and the plasma region, respectively. The horizontal solid black line shows the cross-section plotted in (a).
}
\end{figure}

Oscillations of the driving electric current (see Figure~\ref{fig:04}(a)) create time-varying magnetic field $B_z$ (see Figure~\ref{fig:04}(b)) and, correspondingly, solenoidal electric field with components $E_{sol;x,y}$ (see $E_{sol,y}$ in Figure~\ref{fig:04}(c)). The other electromagnetic field components $B_{x,y}$ and $E_{sol,z}$ are relatively small and are not considered here. The electric field penetrating into the plasma creates oscillating electric current $J_{x,y}$ (see $J_y$ in Figure~\ref{fig:04}(d)). The plasma electrons are energized and produce ionization, the average electron energy and ionization frequency make two oscillations per one period of the driving current, see Figures~\ref{fig:04}(e) and \ref{fig:04}(f).

Examples of instantaneous snapshots of the magnetic field $B_z$, solenoidal electric field $E_{sol;x,y}$, and solenoidal electric current density $J_{sol;x,y}$ are shown in Figures~\ref{fig:05}, \ref{fig:06}, and \ref{fig:07} respectively. The magnetic and electric fields in Figures~\ref{fig:05} and \ref{fig:06} correspond to different times when the magnetic and electric field are in their maximum. The electric current in Figure~\ref{fig:07} corresponds to the time when the driving electric current is almost zero while the solenoidal electric time is maximal (the same time as in  Figure~\ref{fig:06}). This time is chosen to show rings of electric current with opposite direction which appear inside the plasma at certain times as discussed below. Note that the current density shown in Figure~\ref{fig:04}(d) is the full unmodified electric current density $J_y$ while Figure~\ref{fig:07} represents the solenoidal electric current density obtained with Equations (\ref{eq:20}-\ref{eq:21}) and combined with the driving external current (\ref{eq:39}-\ref{eq:40}).

%
%
\begin{figure}
\includegraphics{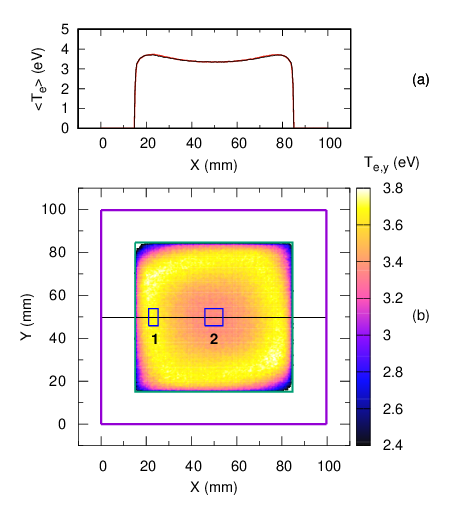}
\caption{\label{fig:09}
(a) Profile of the electron temperature in the $x$ direction $T_{e,x}$ (red curve) and the $y$ direction $T_{e,y}$ (black curve) averaged over time interval $28000\text{~ns}$ - $28095\text{~ns}$ in the cross-section with $y=50\text{~mm}$. (b) The electron temperature in the $y$ direction $T_{e,y}$ versus $x$ and $y$-coordinates at time $t=28094.4\text{~ns}$. In (b), the purple and green squares mark the boundaries of the simulation domain and the plasma region, respectively. Rectangles 1 and 2 mark regions with particles used to calculate the electron velocity distribution function shown in Figure~\ref{fig:13}. The horizontal solid black line shows the cross-section plotted in (a).
}
\end{figure}

The plasma density is maximal in the center of the system and gradually decreases towards the dielectric walls, see Figure~\ref{fig:08}. In the center, the density exceeds $2.5\times 10^{18}\text{~m}^{-3}$. The electron temperature is approximately uniform, in the central region it is slightly smaller than in regions near the plasma boundary (in the skin layer), see Figure~\ref{fig:09}. Except for narrow near-wall regions, the temperature is between $3.3\text{~eV}$ and $3.7\text{~eV}$. The temperatures shown by red and black curves in Figure~\ref{fig:09}(a) are related to the electron motion in the $x$ direction only or the $y$ direction only and are defined as $T_{e,x}=m_e(\langle v_{e,x}^2\rangle-\langle v_{e,x}\rangle^2)$ and $T_{e,y}=m_e(\langle v_{e,y}^2\rangle-\langle v_{e,y}\rangle^2)$, respectively, here $\langle\rangle$ means averaging over particles. The two temperatures show similar behavior, note that the red and black curves in Figure~\ref{fig:09}(a) are virtually indistinguishable. Because of this, only the $T_{e,y}$ is represented by a 2D snapshot in Figure~\ref{fig:09}(b).
%
%
\begin{figure}
\includegraphics{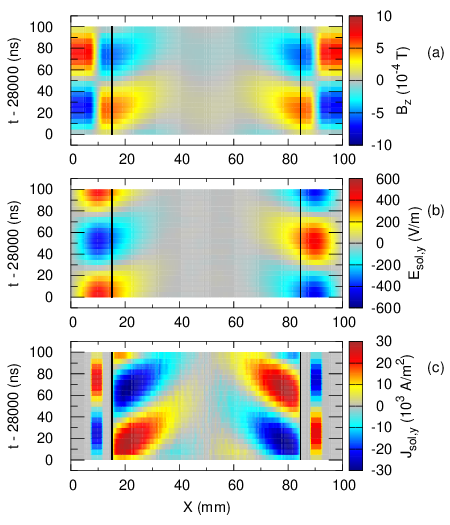}
\caption{\label{fig:10}
Magnetic field $B_z$ (a), solenoidal electric field $E_{sol,y}$ (b), and solenoidal electric current density $J_{sol,y}$ (c) in a cross-section with $y=50\text{~mm}$ versus the $x$-coordinate and time during one period of the driving current starting at $t=28000\text{~ns}$. In (c), the values of the current density in the driving current region are multiplied by $0.1$ to match the scale of the current density in the plasma region. The two vertical black lines show boundaries of the plasma region.
}
\end{figure}

The average frequency of elastic electron-neutral collisions at $t=28000\text{~ns}$ is $\nu_e=1.249\times 10^8\text{~s}^{-1}$, see curve 2 in Figure~\ref{fig:03}(d). In a collisional plasma, the classical skin depth is calculated as
%
%
\begin{equation}\label{eq:42}
    \delta=\dfrac{c}{\omega_e}
           \dfrac{\left(1+{\nu_e^2}/{\omega^2}\right)^{1/4}}
                 {\cos{[\tan^{-1}(\nu_e/\omega)/2]}}~,
\end{equation}
where $\omega_e$ is the electron plasma frequency and $\omega$ is the electromagnetic wave frequency.\cite{KolobovPSST1997} Since the density and temperature are nonuniform, below the skin depth is estimated using the electron density at $x=20\text{~mm}$ where the electron temperature has a maximum. With $\nu_e$ as above, $\omega=2\pi\times 10^7\text{~s}^{-1}$, and $\omega_e=6.176\times 10^{10}\text{~s}^{-1}$ calculated for the density of $1.2\times 10^{18}\text{~m}^{-3}$ at this point, see Figure~\ref{fig:08}(a), the classical skin depth is $\delta=8.5\text{~mm}$. For the electron temperature of $3.7\text{~eV}$, the thermal velocity is $v_{th}=1.14\times 10^6\text{~m/s}$ and the effective electron mean free path is $v_{th}/(\nu_e^2+\omega^2)^{1/2}=8.2\text{~mm}$. Since $\delta\approx v_{th}/(\nu_e^2+\omega^2)^{1/2}$, the skin effect is in a transitional mode between the classical and anomalous modes.\cite{GodyakPRE2001}

The actual width [in the direction normal to the plasma region boundary] of the area in plasma where the electromagnetic field is significant is about $25\text{~mm}$. In the center of the plasma, there is a region about $20\text{~mm}$ wide where the magnetic field amplitude is an order of magnitude less than the amplitude at the plasma boundary, see Figures~\ref{fig:05} and \ref{fig:10}(a). A similar region exists for the solenoidal electric field, see Figures~\ref{fig:06} and \ref{fig:10}(b). The solenoidal electric field there is small because the oscillating magnetic field is weak. The magnetic field of the driving electric current is screened by the electric current in the plasma. Interestingly, the plasma current cancelling the magnetic field of the driving electric current at its maximum is generated earlier while the solenoidal electric field is the strongest.

The solenoidal electric current in the plasma flows along the plasma boundary forming a ring structure, see Figure~\ref{fig:07}. The current induced near the plasma region boundary propagates towards the plasma center (i.e., the ring of current shrinks) with the speed about $5\times 10^5\text{~m/s}$, see Figure~\ref{fig:10}(c). Note that this speed is comparable to the electron thermal velocity used for the estimate of the skin depth above. The shrinking lasts for about half of the driving current period before the amplitude of the current in the ring reduces significantly due to collisions. While shrinking, the current travels about $25\text{~mm}$ in the radial direction, or about 3 effective electron mean free path lengths mentioned above. Before one current ring completely dissipates, another ring with the opposite current forms at the plasma edge. The two rings of plasma electric current with opposite directions co-exist when the driving electric current is minimal, see Figure~\ref{fig:10}(c) for times $0\text{~ns}$, $50\text{~ns}$, and $100\text{~ns}$. Such rings are shown in Figure~\ref{fig:07}. When the driving electric current is maximal, there is only one ring of the plasma electric current directed against the driving current, see Figure~\ref{fig:10}(c) for times $25\text{~ns}$ and $75\text{~ns}$. The strong plasma current cancels most of the magnetic field of the driving current in the central region.

%
%
\begin{figure}
\includegraphics{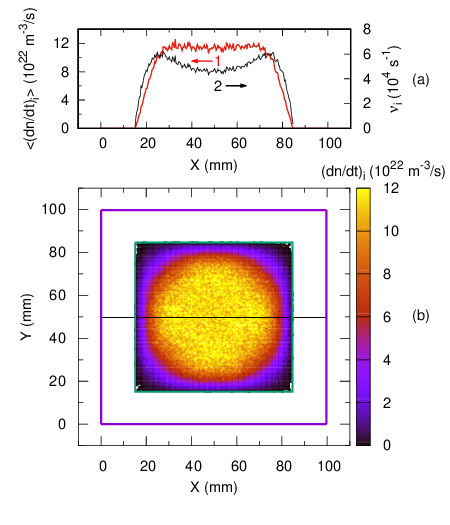}
\caption{\label{fig:11}
(a) Curve 1 (red) shows the profile of the ionization rate averaged over time interval $27000\text{~ns}$ - $28000\text{~ns}$ in a cross-section with $y=50\text{~mm}$, it corresponds to the left vertical coordinate axis (note the red arrow pointing left); curve 2 (black) is the ionization collision frequency, it corresponds to the right vertical coordinate axis (note the black arrow pointing right). (b) The ionization rate averaged over time interval $27900\text{~ns}$ - $28000\text{~ns}$ versus the $x$ and $y$-coordinates. In (b), the purple and green squares mark the boundaries of the simulation domain and the plasma region, respectively. The horizontal solid black line shows the cross-section plotted in (a).
}
\end{figure}

%
%
\begin{figure}
\includegraphics{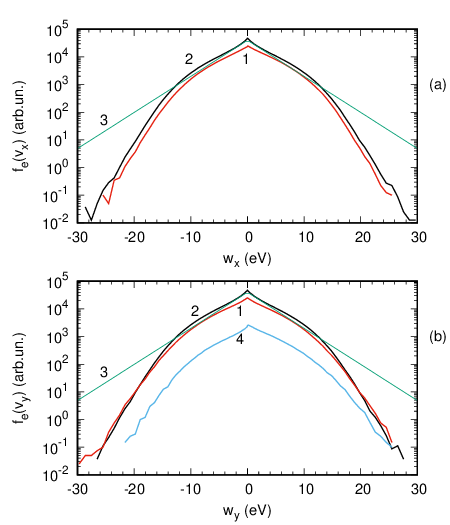}
\caption{\label{fig:12}
Electron velocity distribution functions over the velocity along the $x$ direction (a) and $y$ direction (b) plotted versus the electron energy of motion along the corresponding direction, $w_x=m_ev_{e,x}^2/2$ and $w_y=m_ev_{e,y}^2/2$, respectively. Negative values of the energy correspond to particles moving in the negative direction. Curves 1 (red) and 2 (black) are calculated using particles within regions 1 and 2 shown in Figure~\ref{fig:09}(b) and averaged over time interval $28000\text{~ns}$ - $28095\text{~ns}$. Curve 3(green) is a Maxwellian velocity distribution for temperature 3.33 eV. In (b), curve 4 (blue) is the instantaneous velocity distribution in region 1 at time $28064.5\text{~ns}$, for clarity this curve is shifted downward by one order of magnitude.
}
\end{figure}

The ionization rate is maximal and approximately uniform in a wide region in the center of the plasma, see Figure~\ref{fig:11}(b) and curve 1 in Figure~\ref{fig:11}(a). The ionization collision frequency, however, is maximal in the skin layer and has a minimum in the center, see curve 2 in Figure~\ref{fig:11}(a). This curve is obtained as a ratio of the ionization rate of curve 1 in Figure~\ref{fig:11}(a) and the electron density curve in Figure~\ref{fig:08}(a). The ionization frequency profile is qualitatively similar to the electron temperature profile but the relative difference between the near-wall peaks and the center is noticeably larger for the ionization frequency, compare curve 2 in Figure~\ref{fig:11}(a) and the curves in Figure~\ref{fig:09}(a). The reason for this is the non-Maxwellian electron velocity distribution function (EVDF).

The EVDF is depleted for energies above 16 eV which is close to the ionization threshold for Argon, compare curves 1 and 2 with the Maxwellian EVDF curve 3 in Figure~\ref{fig:12}. In the skin layer, the EVDF over velocity parallel to the direction of the solenoidal electric field (e.g. curve 1 (red) in Figure~\ref{fig:12}(b)) has the number of particles with energy above the ionization potential [relative to the number of particles forming the EVDF maximum] larger than the same EVDF in the middle of the plasma region (curve 2 (black) in Figure~\ref{fig:12}(b)). The EVDFs over velocity normal to the plasma boundary are similar to each other (i.e. without relative difference in the number of electrons for all energies) in the skin layer and in the center of the plasma region, compare curves 1 and 2 in Figure~\ref{fig:12}(a).

In the center of the plasma, the EVDF is isotropic: curve 2 in Figure~\ref{fig:12}(a) is very close to curve 2 in Figure~\ref{fig:12}(b). In the skin layer, the EVDF is not isotropic: curves 1 in Figures~\ref{fig:12}(a) and \ref{fig:12}(b) are different which is expected since the solenoidal electric field contributes mostly to the energy of electron motion in the direction parallel to the field. One may notice that the difference does not affect the values of temperatures $T_{e,x}$ and $T_{e,y}$ which are very close to each other, compare red and black curves in Figure~\ref{fig:09}(a). The reason is that the EVDFs shown by curves 1 and 2 in Figure~\ref{fig:12} are averaged over one period of the driving electric current for the sake of improved statistics in depleted segments. Instantaneous EVDFs over $v_y$ in the skin layer in the region used to calculate curve 1 are asymmetric because of the induced electric current (i.e. electron flow) along the $y$-direction, an example of such an EVDF is curve 4 in Figure~\ref{fig:12}(b). Temperatures shown in Figure~\ref{fig:09} are obtained from instantaneous EVDFs, the asymmetry is accounted for via the flow velocity in the definition of the temperature given above. The averaging of EVDF over the driving current period transformed the asymmetry of multiple EVDFs into the extended width of symmetric curve 1 in  Figure~\ref{fig:12}(b). The temperature of time-averaged EVDF may be different from time-averaged temperatures obtained from instantaneous EVDFs.

%
%
\begin{figure}
\includegraphics{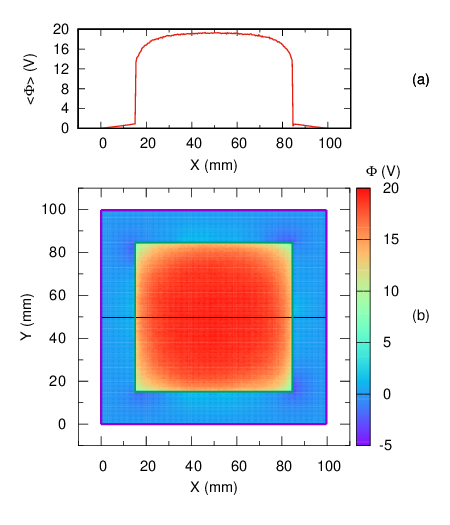}
\caption{\label{fig:13}
(a) Profile of the electrostatic potential averaged over time interval $28000\text{~ns}$ - $28095\text{~ns}$ in the cross-section with $y=50\text{~mm}$. (b) The potential versus the $x$ and $y$-coordinates at time $t=28095\text{~ns}$. In (b), the purple and green squares mark the boundaries of the simulation domain and the plasma region, respectively. The horizontal solid black line shows the cross-section plotted in (a).
}
\end{figure}

%
%
\begin{figure}
\includegraphics{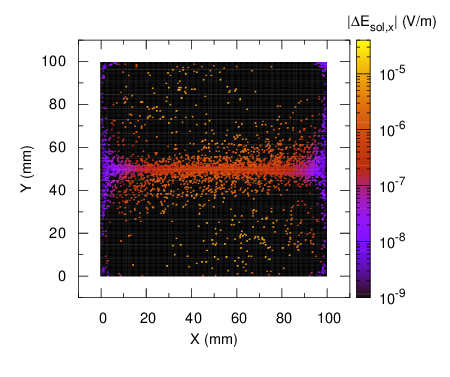}
\caption{\label{fig:14}
The absolute difference between $E_{sol,x}$ calculated with the vorticity method and the SDF method at $t=5\text{~ns}$.
}
\end{figure}

%
%
\begin{figure}
\includegraphics{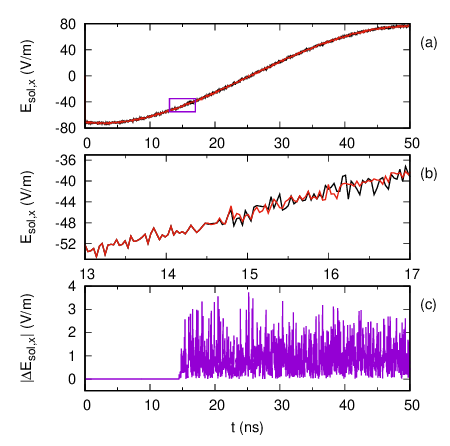}
\caption{\label{fig:15}
(a) The solenoidal electric field $E_{sol,x}$ versus time during half a period of the driving current in a probe with location shown by a black diagonal cross in Figure~\ref{fig:17}. (b) Same as panel (a) around the time when the SDF and vorticity solutions stop being identical. Note that the region shown in (b) corresponds to the purple rectangle in (a). The red and black curves in (a) and (b) correspond to the fields obtained with the SDF and vorticity methods, respectively. (c) The absolute difference between $E_{sol,x}$ calculated with the vorticity method and the SDF method versus time in the probe shown by a black diagonal cross in Figure~\ref{fig:16}.
}
\end{figure}

%
%
\begin{figure}
\includegraphics{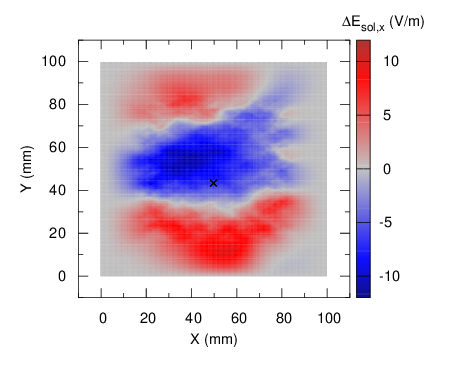}
\caption{\label{fig:16}
The difference between $E_{sol,x}$ calculated with the vorticity method and the SDF method at $t=2999.7\text{~ns}$. The diagonal cross marks location of the probe where the time dependencies shown in Figure~\ref{fig:15} are obtained.
}
\end{figure}

%
%
\begin{figure}
\includegraphics{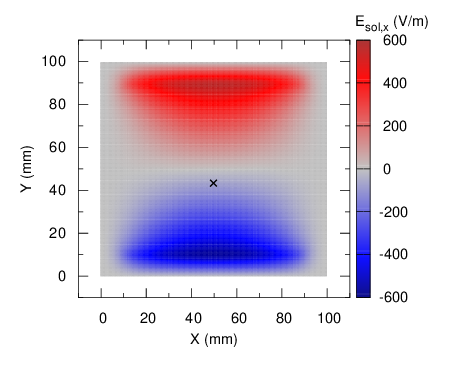}
\caption{\label{fig:17}
The solenoidal electric field $E_{sol,x}$ obtained at $t=2999.7\text{~ns}$ using the vorticity method. The diagonal cross marks location of the probe where the time dependencies shown in Figure~\ref{fig:15} are obtained.
}
\end{figure}

The electrostatic potential is maximal in the center and decays monotonically towards the walls, see Figure~\ref{fig:13}. In the center, the value of the potential averaged over the plasma period is about 19 V. The potential is exactly zero at the metal walls. Charge accumulation on the dielectric boundary results in the potential there being small, usually no more than $\pm 3\text{~V}$, but non-zero. A sharp monotonic drop of the potential across a single cell adjacent to the dielectric wall forms automatically and plays the role of a sheath. This potential difference between the center and the dielectric surface is sufficient to contain electrons with energy of motion in the direction normal to the wall about the threshold ionization energy (16 eV). For the average ionization frequency $5\times 10^4\text{~s}^{-1}$, the mean free path between ionization collisions for an electron with the energy $16\text{~eV}$ is about $47\text{~m}$. This distance is much larger than the size of the plasma. Therefore, the tails of the EVDFs are depleted because energetic electrons escape to the walls, not because they all participate in ionization. On the other hand, for the frequency of elastic electron-neutral collisions of $1.24\times 10^8\text{~s}^{-1}$, the mean free path of an electron with thermal energy $3.5\text{~eV}$ is about $9\text{~mm}$. This means that electrons participate in numerous collisions as they travel between the walls. It makes the EVDF more isotropic, and it also extends the time an energetic particle spends inside the plasma region before escaping at the wall which increases probability of an ionization collision. It is necessary to mention that the simulation is performed without the Coulomb collisions which may be important for such a dense plasma and make the EVDF shape more Maxwellian.

\section{\label{sec:04} Comparison of the vorticity and SDF methods}

The simulation described in Section~\ref{sec:03} is repeated with the SDF method. Initially, the two simulations produce almost identical results. For example, at time $5\text{~ns}$ the absolute difference between $E_{sol,x}$ calculated with the two methods does not exceed $4\times 10^{-5}\text{~V/m}$, see Figure~\ref{fig:14}. Starting at time $14.7\text{~ns}$, the two simulations are no longer identical, compare the black and red curves in Figures~\ref{fig:15}(a) and \ref{fig:15}(b), but the simulations do not diverge from each other (note that the black and red curves in Figure~\ref{fig:15}(a) are very close everywhere). At any time, the difference between the solenoidal electric fields obtained with the two methods does not exceed a few V/m, see Figure~\ref{fig:15}(c) and Figure~\ref{fig:16}. For comparison, the peak magnitude of the field at the time corresponding to Figure~\ref{fig:16} is about $600\text{~V/m}$, see Figure~\ref{fig:17}.

\begin{figure}
\includegraphics{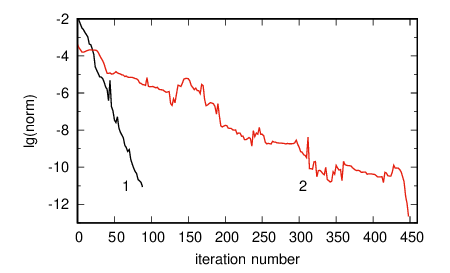}
\caption{\label{fig:18}
The residual norm of the KSPBCGSL solver versus the iteration number during calculation of $E_{sol;x,y}$ at $t=5\text{~ns}$. Curves 1 and 2 correspond to the equation system of the vorticity method and the SDF method, respectively.
}
\end{figure}

\begin{figure}
\includegraphics{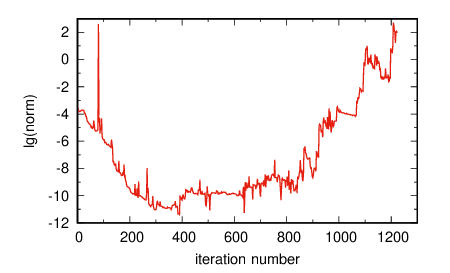}
\caption{\label{fig:19}
The residual norm of the KSPBCGSL solver versus the iteration number during calculation of $E_{sol;x,y}$ with the SDF method at time $t=3012.29\text{~ns}$ when the convergence was not achieved.
}
\end{figure}

The iterative linear equation solvers in both methods are configured to converge when the residual norm reduces by a factor of $10^{-9}$. The vorticity method converges much faster than the SDF method, compare curves 1 and 2 in Figure~\ref{fig:18}. In the vorticity method, the convergence is almost always monotonic and the number of iterations does not change significantly from one time step to another. The SDF method converges non-monotonically (note the multiple jumps of curve 2 in Figure~\ref{fig:18}), the number of iterations may change by hundreds between the two consecutive time steps. Eventually, the SDF method fails to converge at $t=3012.29\text{~ns}$, see Figure~\ref{fig:19}, which stops the simulation.
%
%
%
%


\section{\label{sec:05} Conclusion}

A two-dimensional particle-in-cell code in Cartesian geometry has been developed based on the direct implicit Darwin electromagnetic algorithm described in Ref.~\onlinecite{GibbonsJCP1995}. Significant modifications of the original algorithm have been made. The code uses staggered grids convenient for electromagnetic simulation. The self-consistent magnetic field is calculated using the solenoidal part of the electric current obtained from a Poisson's-like equation. The contribution of collisional scattering is included in calculation of the solenoidal electric fields. The solenoidal electric field in the simulation plane is calculated with a new method using the equation for the vorticity of the solenoidal electric field. The linear system of equations in the vorticity method is solved using a standard iterative solver (PETSc KSPBCGSL solver with the multigrid preconditioner from HYPRE) and the convergence is achieved much faster than for the SDF method of Ref.~\onlinecite{GibbonsJCP1995}.

The code is applied to simulate an inductively coupled plasma system with the driving antenna wrapped around the plasma. Simulation resolves the plane normal to the axis of the system. In such configuration, the components of the solenoidal electric field in the simulation plane are important. A simulation lasting 300 driving current periods has been performed with the vorticity method involved. In simulation, the plasma is sustained by the electromagnetic fields generated by the oscillating driving electric current in the antenna. By the end of simulation, the plasma density increases two orders of magnitude compared to the initial density and acquires a maximum in the center of the system.
%
%
\begin{figure}
\includegraphics{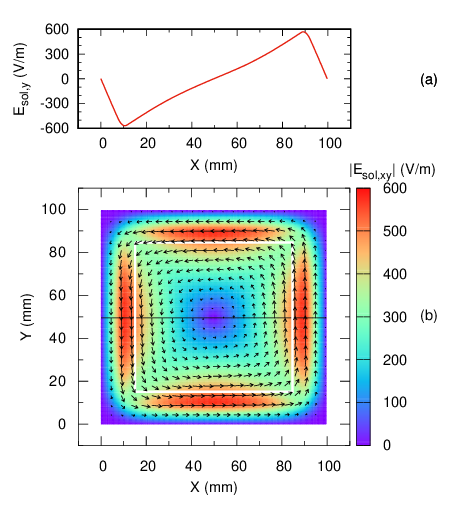}
\caption{\label{fig:20}
Solenoidal electric field at time $t=49.7\text{~ns}$. Panel (a) shows $E_{sol,y}$ versus the $x$-coordinate in the cross-section with $y=50\text{~mm}$. Panel (b) shows the field versus the $x$ and $y$-coordinates. In (b), the colormap shows the absolute value of the field, $|E_{sol,xy}|=(E_{sol,x}^2+E_{sol,y}^2)^{1/2}$, the arrows show the direction of the vector. The cross-section shown in (a) is marked by the horizontal black solid line in (b).
}
\end{figure}

The electron mean free path between collisions with neutrals is small compared to the size of the plasma, but is comparable to the classical skin depth for the achieved values of the plasma density. The skin effect is in a transitional mode between the classical (local) and anomalous (non-local) mode. On one hand, the electric current induced by the oscillating electromagnetic field near the plasma boundary is transported inside the plasma by particles moving normal to the boundary. On the other hand, the current quickly dissipates via collisions. The detailed investigation of the skin effect is out of the scope of the present paper.

The electron velocity distribution is non-Maxwellian, with the high-energy tails strongly depleted for energies above the threshold of confinement by the plasma potential. The electrostatic potential is maximal in the center, the corresponding threshold energy of electron confinement is slightly higher than the ionization threshold. The simulation is in the implicit mode, the size of the numerical grid cell significantly exceeds the electron Debye length. Although the Debye length is not resolved, a drop of potential forms across a single cell adjacent to the wall confining the plasma which plays the role of the plasma sheath.

An attempt to repeat the aforementioned simulation with the SDF method was not successful. Initially both simulations produced identical or very close results, but the field solver using the SDF method failed to converge after 85498 time steps. This demonstrates that the vorticity method is more reliable than the SDF method.
%
%
\begin{figure}
\includegraphics{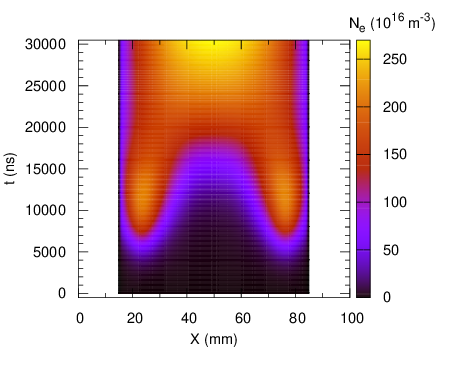}
\caption{\label{fig:21}
Electron density in a cross-section with $y=50\text{~mm}$ versus the $x$-coordinate and time. The cross-section is shown by the solid horizontal black line in Figures~\ref{fig:08}(b) and \ref{fig:22}(b).
}
\end{figure}

%
%
\begin{figure}
\includegraphics{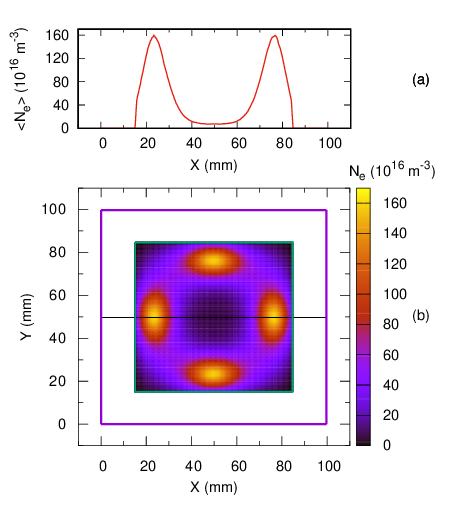}
\caption{\label{fig:22}
Electron density at time $t=8094.4\text{~ns}$: average density profile along the cross-section with $y=50\text{~mm}$ (a) and the density versus the $x$ and $y$-coordinates (b). The values in (a) are averaged over time interval $8000\text{~ns}$ - $8095\text{~ns}$. In (b), the purple and green squares mark the boundaries of the simulation domain and the plasma region, respectively. The cross-section shown in (a) is marked by the horizontal black solid line in (b).
}
\end{figure}

\begin{acknowledgments}

This work at Princeton Plasma Physics Laboratory (PPPL) was supported by the US Department of Energy CRADA agreement between Applied Material Inc. and PPPL.

\end{acknowledgments}

\section*{Data Availability Statement}

The data that support the findings of this study are available from the corresponding author upon reasonable request.

\appendix

\section{\label{app:01} Simulation results during the initial/transitional stage}

The simulation begins with uniform low plasma density and electron temperature. For the initial plasma parameters and the electron-neutral elastic collision frequency of $\nu_e=1.586\times 10^8\text{~s}^{-1}$ the classical skin depth (\ref{eq:42}) is $\delta=106\text{~mm}$. This length exceeds the size of the plasma, therefore, at the initial stage of simulation, the spatial variation of the solenodial electric field is defined not by the skin effect but by the condition of asymmetry relative to the center of the system. Namely, the solenoidal electric field components $E_{sol;x,y}$ are zero in the center and are approximately linear functions of the distance from the center, see Figure~\ref{fig:20}.

\begin{figure}
\includegraphics{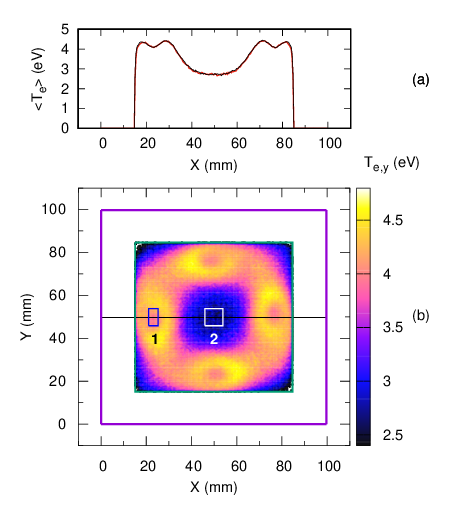}
\caption{\label{fig:23}
(a) Profile of the electron temperature in the $x$ direction $T_{e,x}$ (red curve) and the $y$ direction $T_{e,y}$ (black curve) averaged over time interval $8000\text{~ns}$ - $8095\text{~ns}$ in the cross-section with $y=50\text{~mm}$. (b) The electron temperature in the $y$ direction $T_{e,y}$ versus $x$ and $y$-coordinates at time $t=8094.4\text{~ns}$. In (b), the purple and green squares mark the boundaries of the simulation domain and the plasma region, respectively. Rectangles 1 and 2 mark regions with particles used to calculate the electron velocity distribution function shown in Figure~\ref{fig:26}. The horizontal solid black line shows the cross-section plotted in (a).
}
\end{figure}

\begin{figure}
\includegraphics{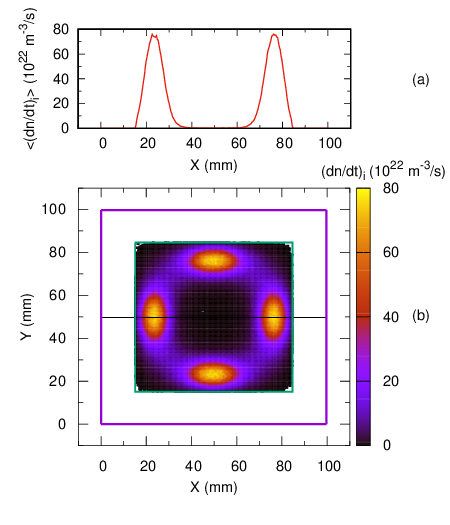}
\caption{\label{fig:24}
(a) Profile of the ionization rate averaged over time interval $7000\text{~ns}$ - $8000\text{~ns}$ in a cross-section with $y=50\text{~mm}$. (b) The ionization rate averaged over time interval $7900\text{~ns}$ - $8000\text{~ns}$ versus the $x$ and $y$-coordinates. In (b), the purple and green squares mark the boundaries of the simulation domain and the plasma region, respectively. The horizontal solid black line shows the cross-section plotted in (a).
}
\end{figure}

Heating of the plasma and, correspondingly, the ionization rate are the strongest in the regions adjacent to the plasma boundary where the solenoidal electric field is the strongest. This causes growth of plasma density within a region of the width of about $20\text{~mm}$ along the plasma boundary, see Figure~\ref{fig:21}. Instantaneous 2D profiles of the electron density and temperature taken at the initial stage show a ring of dense and relatively warm plasma, see Figures~\ref{fig:22} and \ref{fig:23}. The corresponding 2D profiles of the ionization rate also have the characteristic ring structure, see Figure~\ref{fig:24}.

%
%

\begin{figure}
\includegraphics{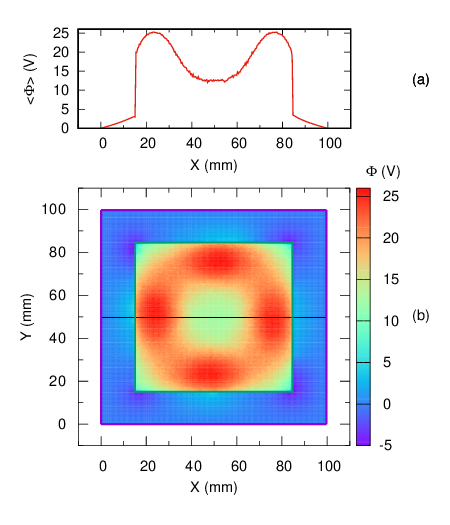}
\caption{\label{fig:25}
(a) Profile of the electrostatic potential averaged over time interval $8000\text{~ns}$ - $8095\text{~ns}$ in the cross-section with $y=50\text{~mm}$. (b) The potential versus the $x$ and $y$-coordinates at time $t=8094.4\text{~ns}$. In (b), the purple and green squares mark the boundaries of the simulation domain and the plasma region, respectively. The horizontal solid black line shows the cross-section plotted in (a).
}
\end{figure}

\begin{figure}
\includegraphics{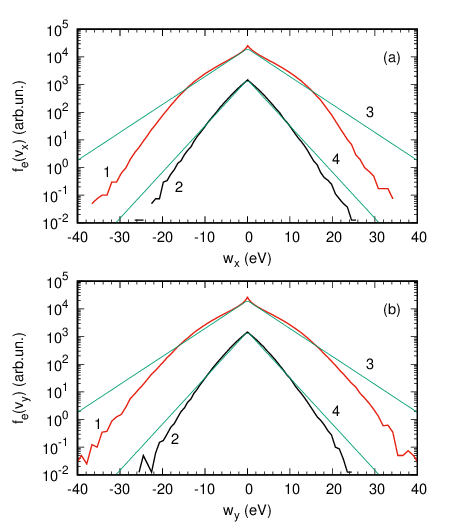}
\caption{\label{fig:26}
Electron velocity distribution functions over the velocity along the $x$ direction (a) and $y$ direction (b) plotted versus the electron energy of motion along the corresponding direction, $w_x=m_ev_{e,x}^2/2$ and $w_y=m_ev_{e,y}^2/2$. Negative values of the energy correspond to particles moving in the negative direction. Curves 1 (red) and 2 (black) are calculated using particles within regions 1 and 2 shown in Figure~\ref{fig:23}(b) and averaged over time interval $8000\text{~ns}$ - $8095\text{~ns}$. Curves 3 and 4 show Maxwellian velocity distributions with temperatures $4.3\text{~eV}$ and $2.6\text{~eV}$, respectively.
}
\end{figure}

In order to maintain plasma quasineutrality, the electrostatic potential at the initial stage has to acquire the ring-shaped profile as well. The potential is maximal near the plasma boundary where the plasma density is maximal and is much lower in the center, see Figure~\ref{fig:25}. The difference between the potential in the maximum and the center can be as large as 10 Volts. Such a profile contains plasma electrons energized by the solenoidal electric field within the near-wall region of the dense plasma. As a result, not only the electron temperature in the center is about $1.7\text{~eV}$ lower than in the near-wall region, but the highest electron energy in the center is about $10\text{~eV}$ smaller than in the near-wall region, compare EVDFs shown by curves 1 and 2 in Figure \ref{fig:26}. This results in the very low ionization rate in the center of the system during the initial stage of simulation.

\begin{figure}
\includegraphics{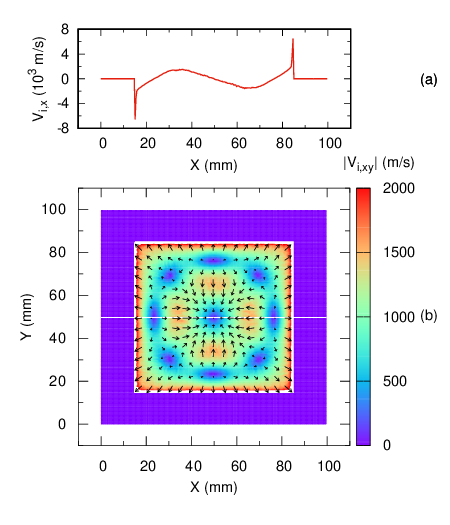}
\caption{\label{fig:27}
(a) Profile of the ion flow velocity $V_{i,x}$  in the cross-section with $y=50\text{~mm}$. (b) The ion flow velocity in the $x$-$y$ plane versus the $x$ and $y$ coordinates. In (b), the colormap shows the absolute value of the velocity, $|V_{i,xy}|=(V_{i,x}^2+V_{i,y}^2)^{1/2}$, the arrows show the direction of the vector. The horizontal solid black line shows the cross-section plotted in (a). All values correspond to time $8094.4\text{~ns}$.
}
\end{figure}

The ions accumulated via ionization gradually accelerate away from the density maximum. This creates ion flows towards the plasma boundary outside the dense plasma ring and towards the center inside the dense plasma ring, see Figure~\ref{fig:27}. The ion flow towards the center is converging which increases the plasma density in the central region. During the first $15000\text{~ns}$ this  mechanism of density increase is much more important than the ionization.

As the density in the center grows, the magnitude of the electrostatic potential barrier gradually decreases which allows more and more energetic electrons to enter the central region and produce ionization there. In addition, the high plasma density reduces the amplitude of the electric field in the plasma, see curve 1 in Figure~\ref{fig:03}(b), which reduces the heating and ionization in the near-wall regions. The density becomes maximal in the center and decaying monotonically towards the boundary of the plasma region at about $22000\text{~ns}$. After this time, the system behaves as described in Section~\ref{sec:03-2}.

%
%

%
%

%
%

%
%

\section{\label{app:02} Selection of numerical parameters}

The direct implicit algorithm may introduce numerical heating or cooling.\cite{CohenJCP1989,HaominPP2023} However, for given grid spatial resolution and time step, there are optimal values of electron density and temperature for which the numerical heating and cooling are negligible.
The optimal set of numerical and plasma parameters is not universal though and depends on details of the simulation setup and the numerical scheme. For example, Ref.~\onlinecite{HaominPP2023} considering 2D electrostatic simulation of a double-periodic collisionless plasma with density $n_{e,0}$ and initial temperature $T_{e,0}$ suggests using $\omega_{e,0}\Delta t \lambda_{e,0}/\Delta x\approx 0.1$. For comparison, Ref.~\onlinecite{CohenJCP1989} considering 1D simulation of a plasma slab expanding into a vacuum half-space concludes that the optimum is when $\omega_{e,0}\Delta t \lambda_{e,0}/\Delta x\approx 0.3$. If in a simulation both the density and temperature of electrons are nonuniform, for a given numerical scheme it is necessary to know, first, how strong the numerical heating or cooling may be when and where the plasma parameters differ from the optimal ones. Second, what can be done to mitigate this effect.

To understand the effect of the direct implicit algorithm on plasmas with different parameters, a set of electrostatic simulations is performed as follows. The electromagnetic part of the algorithm is turned off. The simulation domain is a square filled with plasma only. Periodic boundary conditions are applied at all domain boundaries. The numerical grid is created with the same scale values of $n_{e,0}$, $T_{e,0}$, and $\alpha$ as in Section~\ref{sec:03}, therefore the grid resolution $\Delta x$ is the same. The centered grid has 66 nodes along the $x$ and $y$ directions. There are no collisions with neutrals and no external fields. Initial plasma density and temperature are uniform. The ion mass is 40 AMU, the initial ion temperature is $0.03\text{~eV}$, initial electron and ion velocity distributions are Maxwellian. In the simulation set, the initial plasma density $n_e$ is either $1.6\times 10^{18}\text{~m}^{-3}$ or $1.6\times 10^{16}\text{~m}^{-3}$. The initial electron temperature $T_e$ is $3\text{~eV}$, $4\text{~eV}$, or  $5\text{~eV}$. Parameter $\beta$ defining the time step (\ref{eq:41}) is 8, 9, or 10. A simulation lasts for $50\text{~ns}$. The first set of simulations is performed with the number of particles per cell for the scale density $N_{ppc}=1000$. Note that this value is 8 times lower than the value used in Section~\ref{sec:03}.

In each simulation of the set, the average kinetic electron energy is, approximately, a linear function of time. The slope of this function represents the rate of average kinetic electron energy change with time. The positive or negative rate corresponds to the numerical heating or cooling, respectively. The rates from the first set are assembled in Table~\ref{tab:01}. The numerical cooling is dominant if $\beta=8$. The numerical heating is the dominant process for $\beta=10$. For all values of $\beta$, the rates obtained for the small plasma density are either much larger than the rates obtained for the large density and the same initial temperature, or change the sign. It is necessary to point out that the low density simulations suffer from poor statistics -- there are only 10 particles in a cell for $n_e=1.6\times 10^{16}\text{~m}^{-3}$.
\begin{table}
\caption{\label{tab:01}
The topmost two lines and the leftmost column contain simulation parameters in the set of electrostatic simulations with $N_{ppc}=1000$. The bottom 4 lines, columns 2 to 7, contain the rate of average kinetic electron energy change (in units of $\text{eV}/\mu\text{s}$) for each simulation of the set. The table should be read as follows: for example, the rate value $+1.28\text{~eV}/\mu\text{s}$ in the bottom line of column 2 corresponds to simulation with $n_e=1.6\times 10^{18}\text{~m}^{-3}$, $T_e=3\text{~eV}$, and $\beta=11$, etc.
}
\begin{ruledtabular}
\begin{tabular}{cccccccc}
$n_e (\text{~m}^{-3})$ & \multicolumn{3}{c}{$1.6\times 10^{18}$} & & \multicolumn{3}{c}{$1.6\times 10^{16}$}\\
$T_e$ (eV) & 3.0   & 4.0   & 5.0   & & 3.0   & 4.0   & 5.0 \\
\hline
$\beta=8$  & +0.519 & -0.261 & -1.324 & & -0.260 & -3.697 & -8.956 \\
$\beta=9$  & +1.036 & +0.492 & -0.415 & & +3.270 & -0.491 & -4.067 \\
$\beta=10$ & +1.511 & +1.186 & +0.486 & & +4.816 & +2.064 & -0.837 \\
\end{tabular}
\end{ruledtabular}
\end{table}

Simulations with $\beta=8$ and $\beta=9$ have minimal (in absolute value) energy change rates at $T_e=4\text{~eV}$, see Table~\ref{tab:01}. If the expected range of electron temperatures is between $3\text{~eV}$ and $5\text{~eV}$, one can choose either $\beta=8$ or $\beta=9$ for the actual simulation if the plasma density is large. The value used in Section~\ref{sec:03} is $\beta=9$. Note, however, that with $N_{ppc}=1000$ as in the first set, the simulation is affected significantly by numerical cooling in areas with $T_e>T_{e,0}$ or heating in areas with $T_e<T_{e,0}$ at the transitional stage when the plasma density is low.

Experimenting with the code found that the rate of the numerical heating or cooling reduces for larger values of $N_{ppc}$. Therefore, another simulation set is performed in the same way as the set described above, but with $N_{ppc}=8000$. The results of the second set are assembled in Table~\ref{tab:02}. The energy change rates in the second set are significantly lower than those in the first set. Moreover, rates in the second set obtained for the low density are often smaller than the rates from the first set obtained for the large density (with the same initial electron temperature).
The value $N_{ppc}=8000$ used in the second set and selected for Section~\ref{sec:03} is the largest value which does not push the numerical cost too far for the computational hardware available to the authors.
\begin{table}
\caption{\label{tab:02}
Same as Table~\ref{tab:01}, for the the set of electrostatic simulations with $N_{ppc}=8000$.
}
\begin{ruledtabular}
\begin{tabular}{cccccccc}
$n_e (\text{~m}^{-3})$ & \multicolumn{3}{c}{$1.6\times 10^{18}$} & & \multicolumn{3}{c}{$1.6\times 10^{16}$}\\
$T_e$ (eV) & 3.0    & 4.0    & 5.0    & & 3.0    & 4.0    & 5.0 \\
\hline
$\beta=8$  & +0.058 & -0.044 & -0.191 & & +0.095 & -0.499 & -1.116 \\
$\beta=9$  & +0.130 & +0.054 & -0.071 & & +0.482 & -0.018 & -0.587 \\
$\beta=10$ & +0.187 & +0.145 & +0.049 & & +0.749 & +0.285 & -0.107 \\
\end{tabular}
\end{ruledtabular}
\end{table}

\bibliography{References}

\begin{thebibliography}{24}%
\makeatletter
\providecommand \@ifxundefined [1]{%
 \@ifx{#1\undefined}
}%
\providecommand \@ifnum [1]{%
 \ifnum #1\expandafter \@firstoftwo
 \else \expandafter \@secondoftwo
 \fi
}%
\providecommand \@ifx [1]{%
 \ifx #1\expandafter \@firstoftwo
 \else \expandafter \@secondoftwo
 \fi
}%
\providecommand \natexlab [1]{#1}%
\providecommand \enquote  [1]{``#1''}%
\providecommand \bibnamefont  [1]{#1}%
\providecommand \bibfnamefont [1]{#1}%
\providecommand \citenamefont [1]{#1}%
\providecommand \href@noop [0]{\@secondoftwo}%
\providecommand \href [0]{\begingroup \@sanitize@url \@href}%
\providecommand \@href[1]{\@@startlink{#1}\@@href}%
\providecommand \@@href[1]{\endgroup#1\@@endlink}%
\providecommand \@sanitize@url [0]{\catcode `\\12\catcode `\$12\catcode
  `\&12\catcode `\#12\catcode `\^12\catcode `\_12\catcode `\%12\relax}%
\providecommand \@@startlink[1]{}%
\providecommand \@@endlink[0]{}%
\providecommand \url  [0]{\begingroup\@sanitize@url \@url }%
\providecommand \@url [1]{\endgroup\@href {#1}{\urlprefix }}%
\providecommand \urlprefix  [0]{URL }%
\providecommand \Eprint [0]{\href }%
\providecommand \doibase [0]{http://dx.doi.org/}%
\providecommand \selectlanguage [0]{\@gobble}%
\providecommand \bibinfo  [0]{\@secondoftwo}%
\providecommand \bibfield  [0]{\@secondoftwo}%
\providecommand \translation [1]{[#1]}%
\providecommand \BibitemOpen [0]{}%
\providecommand \bibitemStop [0]{}%
\providecommand \bibitemNoStop [0]{.\EOS\space}%
\providecommand \EOS [0]{\spacefactor3000\relax}%
\providecommand \BibitemShut  [1]{\csname bibitem#1\endcsname}%
\let\auto@bib@innerbib\@empty
\bibitem [{\citenamefont {Birdsall}\ and\ \citenamefont
  {Langdon}(1991)}]{BirdsallBook1991}%
  \BibitemOpen
  \bibfield  {author} {\bibinfo {author} {\bibfnamefont {C.~K.}\ \bibnamefont
  {Birdsall}}\ and\ \bibinfo {author} {\bibfnamefont {A.~B.}\ \bibnamefont
  {Langdon}},\ }\href@noop {} {\emph {\bibinfo {title} {Plasma Physics via
  Computer Simulations}}}\ (\bibinfo  {publisher} {Bristol and Philadelphia :
  IOP Publishing},\ \bibinfo {year} {1991})\BibitemShut {NoStop}%
\bibitem [{\citenamefont {Chen}, \citenamefont {Chacon},\ and\ \citenamefont
  {Barnes}(2011)}]{ChenJCP2011}%
  \BibitemOpen
  \bibfield  {author} {\bibinfo {author} {\bibfnamefont {G.}~\bibnamefont
  {Chen}}, \bibinfo {author} {\bibfnamefont {L.}~\bibnamefont {Chacon}}, \ and\
  \bibinfo {author} {\bibfnamefont {D.~C.}\ \bibnamefont {Barnes}},\ }\bibfield
   {title} {\enquote {\bibinfo {title} {An energy- and charge-conserving,
  implicit, electrostatic particle-in-cell algorithm},}\ }\href@noop {}
  {\bibfield  {journal} {\bibinfo  {journal} {J.\ Comp.\ Phys.}\ }\textbf
  {\bibinfo {volume} {230}},\ \bibinfo {pages} {7018--7036} (\bibinfo {year}
  {2011})}\BibitemShut {NoStop}%
\bibitem [{\citenamefont {Angus}\ \emph {et~al.}(2024)\citenamefont {Angus},
  \citenamefont {Farmer}, \citenamefont {Friedman}, \citenamefont {Geyko},
  \citenamefont {Ghosh}, \citenamefont {Grote}, \citenamefont {Larson},\ and\
  \citenamefont {Link}}]{AngusJCP2024}%
  \BibitemOpen
  \bibfield  {author} {\bibinfo {author} {\bibfnamefont {J.~R.}\ \bibnamefont
  {Angus}}, \bibinfo {author} {\bibfnamefont {W.}~\bibnamefont {Farmer}},
  \bibinfo {author} {\bibfnamefont {A.}~\bibnamefont {Friedman}}, \bibinfo
  {author} {\bibfnamefont {V.}~\bibnamefont {Geyko}}, \bibinfo {author}
  {\bibfnamefont {D.}~\bibnamefont {Ghosh}}, \bibinfo {author} {\bibfnamefont
  {D.}~\bibnamefont {Grote}}, \bibinfo {author} {\bibfnamefont
  {D.}~\bibnamefont {Larson}}, \ and\ \bibinfo {author} {\bibfnamefont
  {A.}~\bibnamefont {Link}},\ }\bibfield  {title} {\enquote {\bibinfo {title}
  {An implicit particle code with exact energy and charge conservation for
  studies of dense plasmas in axisymmetric geometries},}\ }\href {\doibase
  https://doi.org/10.1016/j.jcp.2024.113427} {\bibfield  {journal} {\bibinfo
  {journal} {Journal of Computational Physics}\ }\textbf {\bibinfo {volume}
  {519}},\ \bibinfo {pages} {113427} (\bibinfo {year} {2024})}\BibitemShut
  {NoStop}%
\bibitem [{\citenamefont {Langdon}, \citenamefont {Cohen},\ and\ \citenamefont
  {Friedman}(1983)}]{LangdonJCP1983}%
  \BibitemOpen
  \bibfield  {author} {\bibinfo {author} {\bibfnamefont {A.~B.}\ \bibnamefont
  {Langdon}}, \bibinfo {author} {\bibfnamefont {B.~I.}\ \bibnamefont {Cohen}},
  \ and\ \bibinfo {author} {\bibfnamefont {A.}~\bibnamefont {Friedman}},\
  }\bibfield  {title} {\enquote {\bibinfo {title} {Direct implicit large
  time-step particle simulation of plasmas},}\ }\href@noop {} {\bibfield
  {journal} {\bibinfo  {journal} {J.\ Comput.\ Phys.}\ }\textbf {\bibinfo
  {volume} {51}},\ \bibinfo {pages} {107--138} (\bibinfo {year}
  {1983})}\BibitemShut {NoStop}%
\bibitem [{\citenamefont {Cohen}, \citenamefont {Langdon},\ and\ \citenamefont
  {Hewett}(1989)}]{CohenJCP1989}%
  \BibitemOpen
  \bibfield  {author} {\bibinfo {author} {\bibfnamefont {B.~I.}\ \bibnamefont
  {Cohen}}, \bibinfo {author} {\bibfnamefont {A.~B.}\ \bibnamefont {Langdon}},
  \ and\ \bibinfo {author} {\bibfnamefont {D.~W.}\ \bibnamefont {Hewett}},\
  }\bibfield  {title} {\enquote {\bibinfo {title} {Performance and optimization
  of direct implicit particle simulations},}\ }\href@noop {} {\bibfield
  {journal} {\bibinfo  {journal} {J.\ Comput.\ Phys.}\ }\textbf {\bibinfo
  {volume} {81}},\ \bibinfo {pages} {151--168} (\bibinfo {year}
  {1989})}\BibitemShut {NoStop}%
\bibitem [{\citenamefont {Yee}(1966)}]{YeeIEEETAP1966}%
  \BibitemOpen
  \bibfield  {author} {\bibinfo {author} {\bibfnamefont {K.}~\bibnamefont
  {Yee}},\ }\bibfield  {title} {\enquote {\bibinfo {title} {Numerical solution
  of initial boundary value problems involving maxwell’s equations in
  isotropic media},}\ }\href@noop {} {\bibfield  {journal} {\bibinfo  {journal}
  {IEEE Transactions on Antennas and Propagation}\ }\textbf {\bibinfo {volume}
  {14}},\ \bibinfo {pages} {302--307} (\bibinfo {year} {1966})}\BibitemShut
  {NoStop}%
\bibitem [{\citenamefont {Mattei}\ \emph {et~al.}(2017)\citenamefont {Mattei},
  \citenamefont {Nishida}, \citenamefont {Onai}, \citenamefont {Lettry},
  \citenamefont {Tran},\ and\ \citenamefont {Hatayama}}]{MatteiJCP2017}%
  \BibitemOpen
  \bibfield  {author} {\bibinfo {author} {\bibfnamefont {S.}~\bibnamefont
  {Mattei}}, \bibinfo {author} {\bibfnamefont {K.}~\bibnamefont {Nishida}},
  \bibinfo {author} {\bibfnamefont {M.}~\bibnamefont {Onai}}, \bibinfo {author}
  {\bibfnamefont {J.}~\bibnamefont {Lettry}}, \bibinfo {author} {\bibfnamefont
  {M.~Q.}\ \bibnamefont {Tran}}, \ and\ \bibinfo {author} {\bibfnamefont
  {A.}~\bibnamefont {Hatayama}},\ }\bibfield  {title} {\enquote {\bibinfo
  {title} {A fully-implicit particle-in-cell monte carlo collision code for the
  simulation of inductively coupled plasmas},}\ }\href {\doibase
  10.1016/j.jcp.2017.09.015} {\bibfield  {journal} {\bibinfo  {journal} {J.\
  Comp.\ Phys.}\ }\textbf {\bibinfo {volume} {350}},\ \bibinfo {pages}
  {891--906} (\bibinfo {year} {2017})}\BibitemShut {NoStop}%
\bibitem [{\citenamefont {Darwin}(1920)}]{DarwinPM1920}%
  \BibitemOpen
  \bibfield  {author} {\bibinfo {author} {\bibfnamefont {C.~G.}\ \bibnamefont
  {Darwin}},\ }\bibfield  {title} {\enquote {\bibinfo {title} {The dynamical
  motions of charged particles},}\ }\href {\doibase 10.1080/14786440508636066}
  {\bibfield  {journal} {\bibinfo  {journal} {The London, Edinburgh, and Dublin
  Philosophical Magazine and Journal of Science}\ }\textbf {\bibinfo {volume}
  {39}},\ \bibinfo {pages} {537--551} (\bibinfo {year} {1920})}\BibitemShut
  {NoStop}%
\bibitem [{\citenamefont {Nielson}\ and\ \citenamefont
  {Lewis}(1976)}]{NielsonMCP1976}%
  \BibitemOpen
  \bibfield  {author} {\bibinfo {author} {\bibfnamefont {C.~W.}\ \bibnamefont
  {Nielson}}\ and\ \bibinfo {author} {\bibfnamefont {H.~R.}\ \bibnamefont
  {Lewis}},\ }\bibfield  {title} {\enquote {\bibinfo {title} {Particle-code
  models in the nonradiative limit},}\ }\href {\doibase
  10.1016/B978-0-12-460816-0.50015-4} {\bibfield  {journal} {\bibinfo
  {journal} {Methods in Computational Physics: Advances in Research and
  Applications}\ }\textbf {\bibinfo {volume} {16}},\ \bibinfo {pages}
  {367--388} (\bibinfo {year} {1976})}\BibitemShut {NoStop}%
\bibitem [{\citenamefont {Lee}\ \emph {et~al.}(2005)\citenamefont {Lee},
  \citenamefont {Davidson}, \citenamefont {Startsev},\ and\ \citenamefont
  {Qin}}]{LeeNIMPRA2005}%
  \BibitemOpen
  \bibfield  {author} {\bibinfo {author} {\bibfnamefont {W.}~\bibnamefont
  {Lee}}, \bibinfo {author} {\bibfnamefont {R.~C.}\ \bibnamefont {Davidson}},
  \bibinfo {author} {\bibfnamefont {E.~A.}\ \bibnamefont {Startsev}}, \ and\
  \bibinfo {author} {\bibfnamefont {H.}~\bibnamefont {Qin}},\ }\bibfield
  {title} {\enquote {\bibinfo {title} {The electromagnetic darwin model for
  intense charged particle beams},}\ }\href {\doibase
  10.1016/j.nima.2005.01.233} {\bibfield  {journal} {\bibinfo  {journal}
  {Nuclear Instruments and Methods in Physics Research A}\ }\textbf {\bibinfo
  {volume} {544}},\ \bibinfo {pages} {353--359} (\bibinfo {year}
  {2005})}\BibitemShut {NoStop}%
\bibitem [{\citenamefont {Eremin}\ \emph {et~al.}(2013)\citenamefont {Eremin},
  \citenamefont {Hemke}, \citenamefont {Brinkmann},\ and\ \citenamefont
  {Mussenbrock}}]{EreminJPD2013}%
  \BibitemOpen
  \bibfield  {author} {\bibinfo {author} {\bibfnamefont {D.}~\bibnamefont
  {Eremin}}, \bibinfo {author} {\bibfnamefont {T.}~\bibnamefont {Hemke}},
  \bibinfo {author} {\bibfnamefont {R.~P.}\ \bibnamefont {Brinkmann}}, \ and\
  \bibinfo {author} {\bibfnamefont {T.}~\bibnamefont {Mussenbrock}},\
  }\bibfield  {title} {\enquote {\bibinfo {title} {Simulations of
  electromagnetic effects in high-frequency capacitively coupled discharges
  using the darwin approximation},}\ }\href {\doibase
  10.1088/0022-3727/46/8/084017} {\bibfield  {journal} {\bibinfo  {journal}
  {Journal of Physics D: Applied Physics}\ }\textbf {\bibinfo {volume} {46}},\
  \bibinfo {pages} {084017} (\bibinfo {year} {2013})}\BibitemShut {NoStop}%
\bibitem [{\citenamefont {Barnes}(2022)}]{BarnesJCP2022}%
  \BibitemOpen
  \bibfield  {author} {\bibinfo {author} {\bibfnamefont {D.~C.}\ \bibnamefont
  {Barnes}},\ }\bibfield  {title} {\enquote {\bibinfo {title} {Time-explicit
  darwin pic algorithm},}\ }\href {\doibase 10.1016/j.jcp.2022.111151}
  {\bibfield  {journal} {\bibinfo  {journal} {J.\ Computational Phys.}\
  }\textbf {\bibinfo {volume} {462}},\ \bibinfo {pages} {111151} (\bibinfo
  {year} {2022})}\BibitemShut {NoStop}%
\bibitem [{\citenamefont {Gibbons}\ and\ \citenamefont
  {Hewett}(1995)}]{GibbonsJCP1995}%
  \BibitemOpen
  \bibfield  {author} {\bibinfo {author} {\bibfnamefont {M.~R.}\ \bibnamefont
  {Gibbons}}\ and\ \bibinfo {author} {\bibfnamefont {D.~W.}\ \bibnamefont
  {Hewett}},\ }\bibfield  {title} {\enquote {\bibinfo {title} {The darwin
  direct implicit particle-in-cell (dadipic) method for simulation of low
  frequency plasma phenomena},}\ }\href@noop {} {\bibfield  {journal} {\bibinfo
   {journal} {J.\ Comput.\ Phys.}\ }\textbf {\bibinfo {volume} {120}},\
  \bibinfo {pages} {231--247} (\bibinfo {year} {1995})}\BibitemShut {NoStop}%
\bibitem [{\citenamefont {Hewett}, \citenamefont {Larson},\ and\ \citenamefont
  {Doss}(1992)}]{HewettJCP1992}%
  \BibitemOpen
  \bibfield  {author} {\bibinfo {author} {\bibfnamefont {D.~W.}\ \bibnamefont
  {Hewett}}, \bibinfo {author} {\bibfnamefont {D.~J.}\ \bibnamefont {Larson}},
  \ and\ \bibinfo {author} {\bibfnamefont {S.}~\bibnamefont {Doss}},\
  }\bibfield  {title} {\enquote {\bibinfo {title} {Solution of simulataneous
  partial differential equations using dynamic adi: Solution of the streamlined
  darwin field equations},}\ }\href@noop {} {\bibfield  {journal} {\bibinfo
  {journal} {J.\ Comput.\ Phys.}\ }\textbf {\bibinfo {volume} {101}},\ \bibinfo
  {pages} {11--24} (\bibinfo {year} {1992})}\BibitemShut {NoStop}%
\bibitem [{\citenamefont {Balay}\ \emph {et~al.}(2023)\citenamefont {Balay},
  \citenamefont {Abhyankar}, \citenamefont {Adams}, \citenamefont {Benson},
  \citenamefont {Brown}, \citenamefont {Brune}, \citenamefont {Buschelman},
  \citenamefont {Constantinescu}, \citenamefont {Dalcin}, \citenamefont
  {Dener}, \citenamefont {Eijkhout}, \citenamefont {Faibussowitsch},
  \citenamefont {Gropp}, \citenamefont {Hapla}, \citenamefont {Isaac},
  \citenamefont {Jolivet}, \citenamefont {Karpeev}, \citenamefont {Kaushik},
  \citenamefont {Knepley}, \citenamefont {Kong}, \citenamefont {Kruger},
  \citenamefont {May}, \citenamefont {McInnes}, \citenamefont {Mills},
  \citenamefont {Mitchell}, \citenamefont {Munson}, \citenamefont {Roman},
  \citenamefont {Rupp}, \citenamefont {Sanan}, \citenamefont {Sarich},
  \citenamefont {Smith}, \citenamefont {Zampini}, \citenamefont {Zhang},
  \citenamefont {Zhang},\ and\ \citenamefont {Zhang}}]{petsc-web-page}%
  \BibitemOpen
  \bibfield  {author} {\bibinfo {author} {\bibfnamefont {S.}~\bibnamefont
  {Balay}}, \bibinfo {author} {\bibfnamefont {S.}~\bibnamefont {Abhyankar}},
  \bibinfo {author} {\bibfnamefont {M.~F.}\ \bibnamefont {Adams}}, \bibinfo
  {author} {\bibfnamefont {S.}~\bibnamefont {Benson}}, \bibinfo {author}
  {\bibfnamefont {J.}~\bibnamefont {Brown}}, \bibinfo {author} {\bibfnamefont
  {P.}~\bibnamefont {Brune}}, \bibinfo {author} {\bibfnamefont
  {K.}~\bibnamefont {Buschelman}}, \bibinfo {author} {\bibfnamefont {E.~M.}\
  \bibnamefont {Constantinescu}}, \bibinfo {author} {\bibfnamefont
  {L.}~\bibnamefont {Dalcin}}, \bibinfo {author} {\bibfnamefont
  {A.}~\bibnamefont {Dener}}, \bibinfo {author} {\bibfnamefont
  {V.}~\bibnamefont {Eijkhout}}, \bibinfo {author} {\bibfnamefont
  {J.}~\bibnamefont {Faibussowitsch}}, \bibinfo {author} {\bibfnamefont
  {W.~D.}\ \bibnamefont {Gropp}}, \bibinfo {author} {\bibfnamefont
  {V.}~\bibnamefont {Hapla}}, \bibinfo {author} {\bibfnamefont
  {T.}~\bibnamefont {Isaac}}, \bibinfo {author} {\bibfnamefont
  {P.}~\bibnamefont {Jolivet}}, \bibinfo {author} {\bibfnamefont
  {D.}~\bibnamefont {Karpeev}}, \bibinfo {author} {\bibfnamefont
  {D.}~\bibnamefont {Kaushik}}, \bibinfo {author} {\bibfnamefont {M.~G.}\
  \bibnamefont {Knepley}}, \bibinfo {author} {\bibfnamefont {F.}~\bibnamefont
  {Kong}}, \bibinfo {author} {\bibfnamefont {S.}~\bibnamefont {Kruger}},
  \bibinfo {author} {\bibfnamefont {D.~A.}\ \bibnamefont {May}}, \bibinfo
  {author} {\bibfnamefont {L.~C.}\ \bibnamefont {McInnes}}, \bibinfo {author}
  {\bibfnamefont {R.~T.}\ \bibnamefont {Mills}}, \bibinfo {author}
  {\bibfnamefont {L.}~\bibnamefont {Mitchell}}, \bibinfo {author}
  {\bibfnamefont {T.}~\bibnamefont {Munson}}, \bibinfo {author} {\bibfnamefont
  {J.~E.}\ \bibnamefont {Roman}}, \bibinfo {author} {\bibfnamefont
  {K.}~\bibnamefont {Rupp}}, \bibinfo {author} {\bibfnamefont {P.}~\bibnamefont
  {Sanan}}, \bibinfo {author} {\bibfnamefont {J.}~\bibnamefont {Sarich}},
  \bibinfo {author} {\bibfnamefont {B.~F.}\ \bibnamefont {Smith}}, \bibinfo
  {author} {\bibfnamefont {S.}~\bibnamefont {Zampini}}, \bibinfo {author}
  {\bibfnamefont {H.}~\bibnamefont {Zhang}}, \bibinfo {author} {\bibfnamefont
  {H.}~\bibnamefont {Zhang}}, \ and\ \bibinfo {author} {\bibfnamefont
  {J.}~\bibnamefont {Zhang}},\ }\href {https://petsc.org/} {\enquote {\bibinfo
  {title} {{PETS}c {W}eb page},}\ }\bibinfo {howpublished}
  {\url{https://petsc.org/}} (\bibinfo {year} {2023})\BibitemShut {NoStop}%
\bibitem [{\citenamefont {et~al.}(2023)}]{hypre-web-page}%
  \BibitemOpen
  \bibfield  {author} {\bibinfo {author} {\bibfnamefont {R.~F.}\ \bibnamefont
  {et~al.}},\ }\href {https://github.com//hypre-space//hypre/} {\enquote
  {\bibinfo {title} {Hypre {W}eb page},}\ }\bibinfo {howpublished}
  {\url{https://github.com//hypre-space//hypre/}} (\bibinfo {year}
  {2023})\BibitemShut {NoStop}%
\bibitem [{\citenamefont {Maiorov}(2009)}]{MaiorovPPR2009}%
  \BibitemOpen
  \bibfield  {author} {\bibinfo {author} {\bibfnamefont {S.~A.}\ \bibnamefont
  {Maiorov}},\ }\bibfield  {title} {\enquote {\bibinfo {title} {Ion drift in a
  gas in an external electric field},}\ }\href@noop {} {\bibfield  {journal}
  {\bibinfo  {journal} {Plasma Physics Reports}\ }\textbf {\bibinfo {volume}
  {35}},\ \bibinfo {pages} {802--812} (\bibinfo {year} {2009})}\BibitemShut
  {NoStop}%
\bibitem [{\citenamefont {Sydorenko}(2006)}]{SydorenkoThesis}%
  \BibitemOpen
  \bibfield  {author} {\bibinfo {author} {\bibfnamefont {D.}~\bibnamefont
  {Sydorenko}},\ }\emph {\bibinfo {title} {Particle-in-Cell Simulations of
  Electron Dynamics in Low Pressure Discharges with Magnetic Fields}},\
  \href@noop {} {\bibinfo {type} {Ph.{D}.\ thesis}},\ \bibinfo  {school}
  {University of Saskatchewan} (\bibinfo {year} {2006})\BibitemShut {NoStop}%
\bibitem [{\citenamefont {Sydorenko}\ and\ \citenamefont
  {Khrabrov}(2021)}]{edipic-1d-web-page}%
  \BibitemOpen
  \bibfield  {author} {\bibinfo {author} {\bibfnamefont {D.}~\bibnamefont
  {Sydorenko}}\ and\ \bibinfo {author} {\bibfnamefont {A.}~\bibnamefont
  {Khrabrov}},\ }\href {https://github.com/PrincetonUniversity/EDIPIC}
  {\enquote {\bibinfo {title} {Edipic-1d {W}eb page},}\ }\bibinfo
  {howpublished} {\url{https://github.com/PrincetonUniversity/EDIPIC}}
  (\bibinfo {year} {2021})\BibitemShut {NoStop}%
\bibitem [{\citenamefont {Sydorenko}\ \emph {et~al.}(2021)\citenamefont
  {Sydorenko}, \citenamefont {Khrabrov}, \citenamefont {Villaphana},
  \citenamefont {Powis},\ and\ \citenamefont {Ethier}}]{edipic-2d-web-page}%
  \BibitemOpen
  \bibfield  {author} {\bibinfo {author} {\bibfnamefont {D.}~\bibnamefont
  {Sydorenko}}, \bibinfo {author} {\bibfnamefont {A.}~\bibnamefont {Khrabrov}},
  \bibinfo {author} {\bibfnamefont {W.}~\bibnamefont {Villaphana}}, \bibinfo
  {author} {\bibfnamefont {A.}~\bibnamefont {Powis}}, \ and\ \bibinfo {author}
  {\bibfnamefont {S.}~\bibnamefont {Ethier}},\ }\href
  {https://github.com/PrincetonUniversity/EDIPIC-2D} {\enquote {\bibinfo
  {title} {Edipic-2d {W}eb page},}\ }\bibinfo {howpublished}
  {\url{https://github.com/PrincetonUniversity/EDIPIC-2D}} (\bibinfo {year}
  {2021})\BibitemShut {NoStop}%
\bibitem [{\citenamefont {Startsev}, \citenamefont {Davidson},\ and\
  \citenamefont {Qin}(2007)}]{StartsevPAC2007}%
  \BibitemOpen
  \bibfield  {author} {\bibinfo {author} {\bibfnamefont {E.~A.}\ \bibnamefont
  {Startsev}}, \bibinfo {author} {\bibfnamefont {R.~C.}\ \bibnamefont
  {Davidson}}, \ and\ \bibinfo {author} {\bibfnamefont {H.}~\bibnamefont
  {Qin}},\ }\bibfield  {title} {\enquote {\bibinfo {title} {Numerical studies
  of the electromagnetic weibel instability in intense charged particle beams
  with large temperature anisotropy using the nonlinear best darwin delta-f
  code},}\ }in\ \href {\doibase 10.1109/PAC.2007.4440094} {\emph {\bibinfo
  {booktitle} {2007 IEEE Particle Accelerator Conference (PAC)}}}\ (\bibinfo
  {year} {2007})\ pp.\ \bibinfo {pages} {4297--4299}\BibitemShut {NoStop}%
\bibitem [{\citenamefont {Kolobov}\ and\ \citenamefont
  {Economou}(1997)}]{KolobovPSST1997}%
  \BibitemOpen
  \bibfield  {author} {\bibinfo {author} {\bibfnamefont {V.~I.}\ \bibnamefont
  {Kolobov}}\ and\ \bibinfo {author} {\bibfnamefont {D.~J.}\ \bibnamefont
  {Economou}},\ }\bibfield  {title} {\enquote {\bibinfo {title} {The anomalous
  skin effect in gas discharge plasmas},}\ }\href@noop {} {\bibfield  {journal}
  {\bibinfo  {journal} {Plasma Sources Sci.\ Technol.}\ }\textbf {\bibinfo
  {volume} {6}},\ \bibinfo {pages} {R1--R17} (\bibinfo {year}
  {1997})}\BibitemShut {NoStop}%
\bibitem [{\citenamefont {Godyak}, \citenamefont {Alexandrovich},\ and\
  \citenamefont {Kolobov}(2001)}]{GodyakPRE2001}%
  \BibitemOpen
  \bibfield  {author} {\bibinfo {author} {\bibfnamefont {V.~A.}\ \bibnamefont
  {Godyak}}, \bibinfo {author} {\bibfnamefont {B.~M.}\ \bibnamefont
  {Alexandrovich}}, \ and\ \bibinfo {author} {\bibfnamefont {V.~I.}\
  \bibnamefont {Kolobov}},\ }\bibfield  {title} {\enquote {\bibinfo {title}
  {Lorentz force effects on the electron energy distribution in inductively
  coupled plasmas},}\ }\href@noop {} {\bibfield  {journal} {\bibinfo  {journal}
  {Phys.\ Rev.\ E}\ }\textbf {\bibinfo {volume} {64}},\ \bibinfo {pages}
  {026406} (\bibinfo {year} {2001})}\BibitemShut {NoStop}%
\bibitem [{\citenamefont {Sun}\ \emph {et~al.}(2023)\citenamefont {Sun},
  \citenamefont {Banerjee}, \citenamefont {Sharma}, \citenamefont {Powis},
  \citenamefont {Khrabrov}, \citenamefont {Sydorenko}, \citenamefont {Chen},\
  and\ \citenamefont {Kaganovich}}]{HaominPP2023}%
  \BibitemOpen
  \bibfield  {author} {\bibinfo {author} {\bibfnamefont {H.}~\bibnamefont
  {Sun}}, \bibinfo {author} {\bibfnamefont {S.}~\bibnamefont {Banerjee}},
  \bibinfo {author} {\bibfnamefont {S.}~\bibnamefont {Sharma}}, \bibinfo
  {author} {\bibfnamefont {A.~T.}\ \bibnamefont {Powis}}, \bibinfo {author}
  {\bibfnamefont {A.~V.}\ \bibnamefont {Khrabrov}}, \bibinfo {author}
  {\bibfnamefont {D.}~\bibnamefont {Sydorenko}}, \bibinfo {author}
  {\bibfnamefont {J.}~\bibnamefont {Chen}}, \ and\ \bibinfo {author}
  {\bibfnamefont {I.~D.}\ \bibnamefont {Kaganovich}},\ }\bibfield  {title}
  {\enquote {\bibinfo {title} {{Direct implicit and explicit energy-conserving
  particle-in-cell methods for modeling of capacitively coupled plasma
  devices}},}\ }\href {\doibase 10.1063/5.0160853} {\bibfield  {journal}
  {\bibinfo  {journal} {Physics of Plasmas}\ }\textbf {\bibinfo {volume}
  {30}},\ \bibinfo {pages} {103509} (\bibinfo {year} {2023})}\BibitemShut
  {NoStop}%
\end{thebibliography}%

\end{document}